\documentclass[epj,numbook,nopacs]{svjour}
\usepackage{graphics}
\usepackage{times}
\usepackage{amssymb,amsfonts,amsmath}   
\begin{document}
\title{Irreducible Multiplets of Three-Quark Operators on the Lattice}
\subtitle{Controlling Mixing under Renormalization\thanks{work supported by BMBF}}
\author{Thomas Kaltenbrunner \and Meinulf G\"ockeler \and Andreas Sch\"afer
}                     
%
%
\institute{Institut f\"ur Theoretische Physik, Universit\"at Regensburg, 93040 Regensburg, Germany}
\date{Received: date / Revised version: date}
%
\abstract{
High luminosity accelerators have greatly increased the interest in semi-exclusive and exclusive reactions involving nucleons. The relevant theoretical information is contained in the nucleon wavefunction and can be parametrized by moments of the nucleon distribution amplitudes, which in turn are linked to matrix elements of three-quark operators. These can be calculated from first principles in lattice QCD. However, on the lattice the problems of operator mixing under renormalization are rather involved. In a systematic approach we investigate this issue in depth. Using the spinorial symmetry group of the hypercubic lattice we derive irreducibly transforming three-quark operators, which allow us to control the mixing pattern.
\PACS{
      {PACS-key}{discribing text of that key}   \and
      {PACS-key}{discribing text of that key}
     } 
} 
\maketitle
\section{Introduction}
\label{intro}
In the investigation of the internal nuclear structure, distribution amplitudes play an essential role. Generally, in calculations dealing with exclusive high-energy processes, one can factorize the associated diagrams into hard and soft subprocesses. While a hard subprocess can be evaluated perturbatively and is characteristic for the reaction, the distribution amplitudes describing the nonperturbative soft subprocess are universal \cite{Lepage:1979za,Lepage:1980fj}. Thus, all these computations need the distribution amplitudes as input to produce quantitative results. Presently there exist only model dependent calculations and the QCD sum rule approach \cite{Chernyak:1984bm,King:1986wi,Braun:2006hz}, while the method of choice for the determination of distribution amplitudes from first principles is lattice QCD \cite{Martinelli:1988xs,Gockeler:2007qs}.

After performing an expansion near the lightcone, moments of these nucleon distribution amplitudes are expressed in terms of matrix elements of local three-quark operators that are evaluated between a baryon state and the vacuum and can be computed on the lattice. Apart from isospin symmetrization and color antisymmetrization, these three-quark operators typically look like
\begin{equation}
D_{\mu_1} \dots D_{\mu_m} f_\alpha(x) \cdot D_{\nu_1} \dots D_{\nu_n} g_\beta(x) \cdot D_{\lambda_1} \dots D_{\lambda_l} h_\gamma(x),
\label{el3qOp}
\end{equation}
where $f$, $g$ and $h$ denote the three quark fields with spinor indices $\alpha$, $\beta$ and $\gamma$, respectively, located at some space-time point $x$. 

In order to enable fully quantitative predictions for exclusive baryonic processes a detailed understanding of these three-quark operators is essential. As they pick up radiative corrections and are subject to mixing with other operators, their renormalization is a vital ingredient for any lattice calculation. This paper will focus on a detailed analysis of the operator mixing under renormalization, which is generally constrained by symmetries.

In the Euclidean continuum theory, mixing between operators is restricted by their transformation properties under the symmetry group $O_4$. Operators that transform according to inequivalent irreducible representations of the symmetry group cannot mix under renormalization. However, on the lattice this symmetry group is reduced to its discretized counterpart $H(4)$, which means that in general more operators will participate in the mixing process. In \cite{Goeckeler} a generic study for quark-antiquark operators was performed along these lines. We will modify this approach to deal with the half-integer spin assigned to our three-quark operators. In order to gain control of the renormalization properties, we will construct irreducibly transforming multiplets of three-quark operators with respect to the spinorial hypercubic group $\overline{H(4)}$.

\section{The Symmetry of the Hypercubic Lattice}
\label{sec:1}
\begin{sloppypar}
As the positions of the covariant derivatives within an operator do not affect any of the following arguments, we will assume for ease of notation that all of them act on the last quark field. Given that we will work with linear combinations of the elementary three-quark operators in eq. (\ref{el3qOp}), let us furthermore introduce tensors $T^{(i)}$ that represent their coefficients. Suppressing the color indices and omitting the common space-time coordinate a local three-quark operator then generally looks like:
\end{sloppypar}

\begin{equation}
\mathcal{O}^{(i)} = T^{(i)}_{\alpha \beta \gamma \mu_1 \dots \mu_n} f_\alpha g_\beta D_{\mu_1} \dots D_{\mu_n} h_\gamma.
\label{3qOp}
\end{equation}
The regularized bare operator is related to its renormalized counterpart $\mathcal{O}^{(i),\text{ren}}$ by a renormalization matrix $Z$,
\begin{equation}
\mathcal{O}^{(i),\text{ren}} = Z_{ij} \mathcal{O}^{(j),\text{bare}},
\end{equation}
and mixing under renormalization shows up in non-vanishing off-diagonal elements of $Z$. Given the large number of degrees of freedom (note that $j=1,\dots, 4^{3+n}$), controlling mixing is obviously a complex problem. However, by appropriately choosing the coefficients $T^{(i)}$, it can be achieved that for any given $i$ the number of bare operators contributing with a non-vanishing coefficient $Z_{ij}$ is restricted to a minimum. As mentioned in the introduction, this is done by studying the transformation properties of the given operators under rotations and reflections in space-time. Three-quark operators that do not transform identically to one another do not mix. As any two operators $\mathcal{O}^{(i)}$ and  $\mathcal{O}^{(j)}$ that belong to inequivalent irreducible representations of $\overline{H(4)}$ fulfill this requirement, their related renormalization matrix elements $Z_{ij}$ and $Z_{ji}$ vanish. Once all irreducibly transforming multiplets of three-quark operators are known, the identically transforming operators can be read off. Then $Z$ decomposes into a block diagonal form with one block assigned to each set of identically transforming operators. This greatly facilitates keeping track of the mixing.

\subsection{The Hypercubic Group}
\label{sec:2}
In this section we introduce the symmetry group of the hypercubic lattice (see, e.g., \cite{BGO82}). This so-called hypercubic group $H(4)$ determines, how objects with integer spin behave under transformations of the discretized space-time.

\begin{sloppypar}
In terms of group theory, the hypercubic lattice can be thought of as a set of symmetry transformations of its axes. Let $e_j$, $j=1,\dots,4$ denote unit vectors pointing in the direction of the four canonical axes and let us arrange them as shown in Table~\ref{tab:1}. The symmetry group of a lattice consists of all transformations that leave the symmetry itself untouched, i.e., the lattice looks the same before and after the transformation. There are two classes of operations fulfilling this request. The first one is the interchange of two axes: 
\end{sloppypar}
\begin{equation}
(e_i,-e_i) \leftrightarrow (e_j,-e_j).
\end{equation}
This corresponds to exchanging two rows in the table. As there is a total of four rows, it is readily seen that these operations represent the permutation group with four elements, $S_4$. Inverting an axis is the other symmetry operation one can think of:
\begin{equation}
(e_i,-e_i) \mapsto (-e_i,e_i).
\end{equation}
In the diagram this means to flip the two entries within one row. The corresponding symmetry is $Z_2$. If one again takes into account all four rows, one arrives at ${Z_2}^4$. Working out the commutation relations between the exchange and reflection of axes, an isomorphism between the symmetry group of the hypercubic lattice and a semidirect product ${Z_2}^4 \rtimes S_4$ (wreath product of $Z_2$ and $S_4$) is found. The group order is $4!\cdot 2^4=384$.

Let us now turn to a more abstract approach. The hypercubic group can be defined by six generators $t$, $\gamma$, $I_1$,..., $I_4$ and a set of generating relations \cite{Dai}:
 \begin{align}
   I_i^2&=1,            & I_iI_j&=I_jI_i           &        tI_1&=I_1t,         & \nonumber\\
   tI_2&=I_4t,          & tI_3&=I_2t,              &        tI_4&=I_3t,         & \nonumber\\
   \gamma I_1 &= I_3,   & \gamma I_2&= I_2\gamma,  & \gamma I_4 &= I_4\gamma,   & \nonumber\\
   \gamma^2&= 1,        & t^3&= 1,                 & (t\gamma)^4&=1.            & 
   \label{genH4scal}
 \end{align}
The generators $I_j$ represent inversions whereas $t$ and $\gamma$ stand for a combined reflection and interchange of the axes. Each element $G \in H(4)$ can be expressed as a product of the generators.

For this group there are all in all twenty inequivalent irreducible representations. They are labeled by $\tau^n_k$ with the superscript $n$ giving the dimension of this representation and the optional subscript $k$ counting inequivalent representations of the same dimension, if existent \cite{BGO82}.
Every irreducible representation is uniquely identified by the traces of its representation matrices $\tau^n_k(G)$ for the group elements, called characters $\chi$ of the representation:
\begin{align}
 \chi^n_k(G)=\sum_i \tau^n_k(G)_{ii}.
\end{align}
For us, $\tau^4_1$ is of particular interest because its $4\times4$-matrices describe, how Lorentz vectors such as covariant derivatives transform under the group action.

\begin{table}
 \caption{Symmetries of the hypercubic lattice.}
 \label{tab:1}
 \begin{center}
  \begin{tabular}{|c|c|}
   \hline
   $e_1$ & $-e_1$ \\
   \hline
   $e_2$ & $-e_2$ \\
   \hline
   $e_3$ & $-e_3$ \\
   \hline
   $e_4$ & $-e_4$ \\
   \hline
  \end{tabular}
 \end{center}
\end{table}

\subsection{The Spinorial Hypercubic Group}
\label{sec:3}
As already mentioned in the introduction, we are interested in the transformation properties of objects with half-integer spin. The appropriate symmetry group was studied in \cite{Dai} and is called spinorial hypercubic group $\overline{H(4)}$. In contrast to its non-spinorial counterpart, this group has to contain further features like the phase-factors for the lattice analogue of a full rotation. Therefore the defining relations of $H(4)$ must be modified by twisting them with a set of $Z_2$ factors:
\begin{align}
   I_i^2&=-1,           & I_iI_j&=-I_jI_i          &        tI_1&=I_1t,         & \nonumber\\
   tI_2&=I_4t,          & tI_3&=I_2t,              &        tI_4&=I_3t,         & \nonumber\\
   \gamma I_1 &=-I_3,   & \gamma I_2&=-I_2\gamma,  & \gamma I_4 &=-I_4\gamma,   & \nonumber\\
   \gamma^2&=-1,        & t^3&=-1,                 & (t\gamma)^4&=-1.           & 
   \label{genH4spin}
\end{align}
Beyond the irreducible representations directly inherited from $H(4)$, five more irreducible representations are found. Hence these are ``purely spinorial'' and marked with an underscore beneath their dimension: $\tau^{\underline{4}}_1$, $\tau^{\underline{4}}_2$, $\tau^{\underline{8}}$, $\tau^{\underline{12}}_1$ and $\tau^{\underline{12}}_2$. Here the four-dimensional representation $\tau^{\underline{4}}_1$ is important as it describes the transformation of four-spinors under the group action. By construction the group order of $\overline{H(4)}$ is twice that of $H(4)$.

\section{Construction of Irreducible Three-Quark Operators}
\label{sec:4}
In this section we will explain, how irreducibly transforming three-quark operators can be constructed. Under the group action of $\overline{H(4)}$ an operator is converted into a linear combination of other operators. A major step on the way towards irreducible multiplets is to determine, which operators may be present in these linear combinations.

Knowing the representation matrices for spinors and Lo\-rentz vectors one is in principle able to deduce the transformation of any three-quark operator (\ref{3qOp}) under any group element $G \in \overline{H(4)}$. To this end each spinor and Lorentz index is transformed separately with the representation matrices of $\tau^{\underline{4}}_1$ and $\tau^4_1$, respectively, resulting in the $G$-transformed three-quark operator
\begin{equation}
      \mathcal{O}^{(j),G-\text{transformed}} = G_{ij} \mathcal{O}^{(i)}.
      \label{groupaction}
\end{equation}
However, given the large amount of independent operators $\mathcal{O}^{(i)}$, this would yield transformation matrices $G_{ji}$ of rather unhandy dimension.

As the spinorial hypercubic group is embedded in the symmetry group of the Euclidean continuum $\overline{O_4}$, irreducibly transforming operator multiplets of the latter one form a closed set with respect to the group action of $\overline{H(4)}$. In other words: the $\overline{H(4)}$ representation matrices $G_{ij}$ are block-diagonal with respect to multiplets of three-quark operators transforming irreducibly under the continuum group. When choosing appropriate linear combinations within these multiplets their blocks may however decompose into even smaller blocks under the spinorial hypercubic group, resulting in the desired $\overline{H(4)}$ irreducible representations. Using the symmetry group of the Euclidean continuum thus subdivides the problem of searching for $\overline{H(4)}$ irreducible three-quark operators in the whole operator space into the task of decomposing $\overline{O_4}$ irreducible multiplets. This reduces the dimension of the problem considerably.

We will therefore first derive multiplets of irreducibly transforming three-quark operators in $\overline{O_4}$. In a second step a projector for the decomposition into $\overline{H(4)}$ irreducible operator multiplets is constructed. That fixes our choice for the coefficient tensors $T^{(i)}$ in eq.~(\ref{3qOp}).

\subsection{Irreducibility in $\overline{SO_4}$ and $\overline{O_4}$}
\label{sec:5}
Unless stated otherwise, we will focus on the leading-twist case from now on (i.e., twist $3$). For ease of notation let us write all quark fields with dotted and undotted indices in the chiral Weyl representation (cf., e.g., \cite{Peskin}). Then a four-spinor naturally decomposes into two Weyl-spinors of definite chirality (Table \ref{tab:2}) whose transformation properties are characterized by an $SU(2)$ representation. Analogously, we convert the covariant derivatives to an $SU(2) \times SU(2)$ representation by contracting them with the Pauli matrices $\sigma_\mu$. Then the whole three-quark operator transforms as a direct product of $SU(2)$ representations. Now, there exists a homomorphism that links the irreducible representations of $SU(2) \times SU(2)$ to those of $\overline{SO_4}$:
\begin{align}
 SU(2) \times SU(2) \simeq \overline{SO_4}.
\end{align}
Thus we can deduce irreducibly transforming three-quark operators for the latter group by constructing irreducible representations in $SU(2)\times SU(2)$, which is accomplished by appropriately symmetrizing the $SU(2)$ indices according to the corresponding standard Young tableaux \cite{Urbantke}. In leading-twist this enforces independent total symmetrization of the dotted and undotted indices. 

To be specific, let us assume that a particular combination of quark chiralities is given, i.e., the spinor indices are chosen to be either dotted or undotted. Then an $\overline{SO_4}$ irreducibly transforming multiplet is constructed as follows:
\begin{align}
      f_{\dot a} g^{b} D_{\mu_1} \dots D_{\mu_n} h^{c} \, &\to \, f_{\dot a} g^{b} {(D\sigma)^{d_1}}_{\dot e_1} \dots {(D\sigma)^{d_n}}_{\dot e_n} h^{c} \, \nonumber\\
      &\to \,  f_{\{\dot a} g^{\{b} {(D\sigma)^{d_1}}_{\dot e_1} \dots {(D\sigma)^{d_n}}_{\dot e_n\}} h^{c\}}, \label{SO4_chiral}
\end{align}
with $\{ \dots \}$ denoting the symmetrization of the indices on the same level. Looking at the dotted and undotted indices, which each can take the values zero and one, we immediately read off that this multiplet consists of $(n+3)\cdot(n+2)$ three-quark operators. Operators with other chirality combinations of the quark fields are treated in the same manner, so that the space of elementary three-quark operators (\ref{el3qOp}) decomposes into subspaces of $\overline{SO_4}$ irreducible multiplets.

So far only four-dimensional rotations were taken into account. The link to the full symmetry group of the Euclidean continuum $\overline{O_4}$ is given by reflection operations: let $r$ represent some reflection in four dimensions, then $O_4=SO_4 \cup r SO_4$ \cite{BGO83}, which also holds for the covering groups $\overline{O_4}$ and $\overline{SO_4}$. Therefore an $\overline{O_4}$ irreducible multiplet of three-quark operators can be constructed by combining any of the just deduced $\overline{SO_4}$ irreducible multiplets with its parity partner to a larger one.

In the next subsection we will exploit the fact that in the basis of these $\overline{O_4}$ irreducible multiplets all representation matrices $G_{ij}$ in equation (\ref{groupaction}) are simultaneously block-diagonal. This facilitates the further decomposition into the desired $\overline{H(4)}$ irreducible representations.

\begin{table}
 \caption{Relation of quark fields with dotted and undotted indices to the Weyl representation}
 \label{tab:2}
 \begin{tabular}{c cccc}
  \hline\noalign{\smallskip}
  Weyl representation & $\psi_1$ & $\psi_2$ & $\psi_3$           & $\psi_4$ \\
  \noalign{\smallskip}\hline\noalign{\smallskip}
  (un)dotted indices  & $\Phi^0$ & $\Phi^1$ & $\Sigma_{\dot{0}}$ & $\Sigma_{\dot{1}}$\\
  \noalign{\smallskip}\hline\noalign{\smallskip}
  chirality           & $+$      & $+$      & $-$                & $-$ \\
  \noalign{\smallskip}\hline
 \end{tabular}
\end{table}

\subsection{Irreducibility in $\overline{H(4)}$}
\label{sec:6}
Before we explain how the actual decomposition works, let us find out which $\overline{H(4)}$ irreducible representations may show up. As stated in Sections \ref{sec:2} and \ref{sec:3}, the covariant derivatives and quark fields transform according to $\tau^{4}_1$ and $\tau^{\underline{4}}_1$, respectively. Therefore the three-quark operator in (\ref{el3qOp}) transforms as a direct product of these representations:
\begin{equation}
\tau^{\underline{4}}_1 \otimes \tau^{\underline{4}}_1 \otimes \tau^{4}_1 \otimes \dots \otimes \tau^{4}_1 \otimes \tau^{\underline{4}}_1.
\end{equation}
This product is reducible. Knowing the characters $\chi^\alpha$ for a given irreducible representation $\tau^\alpha$, it can be decomposed with the help of the identity 
\begin{align}
 \tau^\alpha \otimes \tau^\beta & = \sum_\gamma c_\gamma \tau^\gamma,
\end{align}
where
\begin{align}
 c_\gamma = \frac{1}{\vert \overline{H(4)} \vert} &\sum_{G \in \overline{H(4)}}  \chi^\gamma(G)^* \cdot \chi^\alpha(G) \cdot \chi^\beta(G)
\end{align}
and $\vert \overline{H(4)} \vert$ denotes the group order.
Applying this formula iteratively, we derive the following content of $\overline{H(4)}$ irreducible multiplets for three-quark operators with zero to two derivatives (including higher twist):
\begin{align}
&\text{zero derivatives:} &&\tau^{\underline{4}}_1 \otimes \tau^{\underline{4}}_1 \otimes \tau^{\underline{4}}_1 = \nonumber\\
                   & &&\quad\quad 5\tau_1^{\underline{4}} \oplus \tau^{\underline{8}} \oplus 3 \tau_1^{\underline{12}}, \nonumber\\
&\text{one derivative:} &&\tau^{\underline{4}}_1 \otimes \tau^{\underline{4}}_1 \otimes \tau^{4}_1 \otimes \tau^{\underline{4}}_1 = \nonumber\\
            & &&\quad\quad 8\tau_1^{\underline{4}} \oplus 4 \tau^{\underline{8}} \oplus 12 \tau_1^{\underline{12}} \oplus 4 \tau_2^{\underline{12}}, \nonumber\\
&\text{two derivatives:} &&\tau^{\underline{4}}_1 \otimes \tau^{\underline{4}}_1 \otimes \tau^{4}_1 \otimes \tau^{4}_1 \otimes \tau^{\underline{4}}_1= \nonumber\\
& &&\quad\quad 20 \tau_1^{\underline 4} \oplus 3 \tau_2^{\underline 4} \oplus 18 \tau^{\underline 8} \oplus 41 \tau_1^{\underline{12}} \oplus 23 \tau_2^{\underline{12}}.
\end{align}
As expected, only spinorial representations show up after the reduction process.

We now proceed with the decomposition of the $\overline{O_4}$ multiplets from the previous section. Therefore, their $768$ transformation matrices $G_{ij}$ in (\ref{groupaction}) must be known explicitly. To be more precise: the diagonal blocks of these matrices are sufficient.

For every group element this matrix block can be construc\-ted by transforming every quark field and derivative of an operator separately as explained above and writing the result in terms of the original operators. The coefficients involved are identified with the representation matrix elements $G_{ij}$. Then, with the knowledge of the characters of the irreducible representations, a projector is constructed. When applied to an $\overline{O_4}$ multiplet it projects out a usually smaller multiplet that transforms according to the desired irreducible $\overline{H(4)}$ representation (see, e.g., \cite{Chesnut}):
\begin{equation}
P^\alpha=\frac{d_\alpha}{\vert \overline{H(4)} \vert} \sum_{G \in \overline{H(4)}} {\chi^\alpha(G)}^{*} \cdot G.
\label{projector}
\end{equation}
Here $d_\alpha$ denotes the dimension of the irreducible representation to be projected out.

Some $\overline{O_4}$ irreducible multiplets contain several equivalent $\overline{H(4)}$ irreducible representations $\tau^\alpha$. Then, the action of $P^\alpha$ yields a set of three-quark operators that actually contains \hyphenation{smal-ler} smaller multiplets, each closed under the group action on its own and irreducible. To seperate these multiplets a second projector $\tilde P^\alpha_{lk}$ is introduced (see, e.g., \cite{Miller}):
\begin{eqnarray}
\tilde P^\alpha_{lk} = \frac{d_\alpha}{\vert \overline{H(4)} \vert} \sum_{G \in \overline{H(4)}} (G_{lk}^\alpha)^{*} G.
\label{equiv_projector}
\end{eqnarray}
Here $G_{lk}^\alpha$ denotes the $lk$ element of the representation matrix $\tau^\alpha(G)$. Acting with $\tilde P^\alpha_{11}$ on the set in question results in $m$ independent three-quark operators, where $m$ is the multiplicity of the representation $\tau^\alpha$. If we now apply the projectors $\{\tilde P^\alpha_{1j}, \quad j=1, \dots, d_\alpha\}$ to these $m$ operators separately we generate $m$ irreducible multiplets of three-quark operators. That results in the requested separation of the $m$ equivalent irreducible multiplets.

After performing these steps all irreducibly transforming three-quark operators of the spinorial hypercubic group are known.

\section{Three-Quark Operators and Renormalization}
\label{sec:7}
In the previous section we have explained how multiplets of $\overline{H(4)}$ irreducibly transforming three-quark operators of leading twist can be constructed. Starting from different Young tableaux in the $\overline{SO_4}$ case the very same concept applies to higher twist. The results are summarized in Appendix \ref{App:1}. There we give a full list of all leading-twist irreducible three-quark operators with up to two derivatives and all higher-twist operators without derivatives.

In the following we want to discuss the consequences for the mixing properties of the lattice operators under renormalization. It was already stated in Section \ref{sec:1} that mixing is prohibited between three-quark operators belonging to inequivalent irreducible representations. In Table \ref{tab:3} we sort our results according to their representation and mass dimension. Here $\mathcal{O}^{(i)}_j$ denotes the $i$-th operator within the $j$-th $\overline{H(4)}$ irreducible multiplet. Then the above statement means that renormalization only mixes operators within the same row. More precisely: we sort the operators within any multiplet in such a way that their transformation matrices under group action are identical for equivalent representations. Thus, the $i$-th operators in multiplets of equivalent representations transform identically and therefore mix only with each other.

When working with dimensional regularization in the continuum theory, mixing is also forbidden for operators with different mass dimensions, i.e., different columns cannot mix. On the lattice, however, this last statement is not valid anymore. Due to the existence of a dimensionful quantity, namely the lattice spacing $a$, lower-dimensional operators may mix with higher-dimensional ones with coefficients proportional to powers of $1/a$, e.g.:
   \begin{equation}
      \mathcal{O}^{(i),\text{ren}}= Z_{ij} \mathcal{O}^{(j),\text{bare}} + Z' \cdot \frac{1}{a} \cdot  \mathcal{O}^{\text{bare, lower dim}}.
   \end{equation}
In practice it proves difficult to properly extract the renormalization coefficients $Z'$ of mixing lower-dimensional operators. Therefore this situation should be avoided wherever possible. To this end one can try to restrict oneself to those representations that do not possess lower-dimensional counterparts such as $\tau^{\underline{12}}_2$ ($\tau^{\underline{4}}_2$) for three-quark operators with one (two) derivatives.

We can summarize these statements: the $i$-th operator of a multiplet may mix with any $i$-th operator of the same or lower dimension from the same row in Table \ref{tab:3}. All operators within one multiplet share the same renormalization coefficients. Hence it is sufficient to renormalize one operator per multiplet only, e.g., $i=1$.

\begin{table}
 \caption{Irreducibly transforming multiplets of three-quark operators sorted by their mass-dimension.}
 \renewcommand{\arraystretch}{1.7}
 \label{tab:3}
 \begin{tabular}{|c||c|c|c|}
   \hline
 & dimension 9/2 & dimension 11/2 & dimension 13/2\\
 & (0 derivatives) & (1 derivative) & (2 derivatives) \\
  \hline
  \hline
  $\tau_1^{\underline{4}}$ & $ \begin{matrix} \mathcal{O}_{1}^{(i)}, \\ \mathcal{O}_{2}^{(i)}, \quad \mathcal{O}_{3}^{(i)}, \\ \mathcal{O}_{4}^{(i)}, \quad \mathcal{O}_{5}^{(i)} \end{matrix} $  & & $\begin{matrix} \mathcal{O}_{DD1}^{(i)}, \\ \mathcal{O}_{DD2}^{(i)}, \quad \mathcal{O}_{DD3}^{(i)} \end{matrix}$\\
  \hline
  $\tau_2^{\underline{4}}$ &  &  & $\begin{matrix} \mathcal{O}_{DD4}^{(i)}, \\ \mathcal{O}_{DD5}^{(i)}, \quad \mathcal{O}_{DD6}^{(i)} \end{matrix}$\\
  \hline
  $\tau^{\underline{8}}$ & $\mathcal{O}_{6}^{(i)}$ &  $\mathcal{O}_{D1}^{(i)}$ & $\begin{matrix} \mathcal{O}_{DD7}^{(i)},\\ \mathcal{O}_{DD8}^{(i)}, \quad \mathcal{O}_{DD9}^{(i)} \end{matrix}$ \\
  \hline
  $\tau_1^{\underline{12}}$ & $\begin{matrix} \mathcal{O}_{7}^{(i)}, \\ \mathcal{O}_{8}^{(i)}, \quad \mathcal{O}_{9}^{(i)} \end{matrix} $ & $\begin{matrix} \mathcal{O}_{D2}^{(i)},\\ \mathcal{O}_{D3}^{(i)}, \quad \mathcal{O}_{D4}^{(i)} \end{matrix}$  & $\begin{matrix} \mathcal{O}_{DD10}^{(i)},\quad \mathcal{O}_{DD11}^{(i)},\\ \mathcal{O}_{DD12}^{(i)},\quad \mathcal{O}_{DD13}^{(i)} \end{matrix}$ \\
  \hline
  $\tau_2^{\underline{12}}$ &  &  $\begin{matrix} \mathcal{O}_{D5}^{(i)}, \quad \mathcal{O}_{D6}^{(i)},\\ \mathcal{O}_{D7}^{(i)}, \quad \mathcal{O}_{D8}^{(i)} \end{matrix}$ & $\begin{matrix} \mathcal{O}_{DD14}^{(i)},\\ \mathcal{O}_{DD15}^{(i)}, \quad \mathcal{O}_{DD16}^{(i)},\\ \mathcal{O}_{DD17}^{(i)}, \quad \mathcal{O}_{DD18}^{(i)} \end{matrix}$\\
  \hline
\end{tabular}
 \renewcommand{\arraystretch}{1.0}
\end{table}

Recall, however, that without loss of generality we have discussed three-quark operators with all derivatives acting on the last quark. That was possible, because the actual position of the derivative has no influence on the transformation properties and thus on the classification for renormalization. Mixing between operators with merely interchanged position of the derivatives is not prohibited. Hence, it is important to note how the additional operators not listed explicitly in Appendix \ref{App:1} are generated. For a given operator only the position of the derivatives is changed to any quark field without touching the spinor and vector indices quoted. This yields three times as many possibly mixing multiplets in the case of one and six times as many in the case of two derivatives.

Let us introduce a notation that characterizes the operators uniquely. We replace the $D$ in the subscript of a multiplet with an $f$, if the derivative acts on the first quark, a $g$ ($h$) if it acts on the second (third) quark. E.g., the operator $\mathcal{O}_{DD17}^{(4)}$ with derivatives acting on the first and second quark then looks like:
\begin{align}
\mathcal{O}_{fg17}^{(4)} = \frac{5i}{4\sqrt{6}} &\left( \frac{3}{5} {(D\sigma)^{\{0}}_{\{ \dot 0} f^{0} {(D\sigma)^0}_{\dot 0 \}} g^{0}  h^{0 \}} \right. \nonumber\\ 
                          & - {(D\sigma)^{\{1}}_{\{ \dot 0} f^{1} {(D\sigma)^1}_{\dot 0 \}} g^{1}  h^{0 \}} \nonumber\\
                          & \left. - 2 \cdot {(D\sigma)^{\{0} }_{\{ \dot 1} f^{1} {(D\sigma)^0}_{\dot 1 \}} g^{1}   h^{0 \}} \right).
\end{align}
Due to the total symmetrization of the spinor indices an interchange of the two derivatives has no effect on the operator, i.e., $\mathcal{O}_{gf17}^{(4)}=\mathcal{O}_{fg17}^{(4)}$. Thus we can always order the indices $f$, $g$ and $h$ alphabetically leaving only six classes of operators with two derivatives, as mentioned above:
\begin{align}
&\mathcal{O}_{ff\dots}^{(i)},&
&\mathcal{O}_{fg\dots}^{(i)},&
&\mathcal{O}_{fh\dots}^{(i)}, \nonumber\\
&\mathcal{O}_{gg\dots}^{(i)},&
&\mathcal{O}_{gh\dots}^{(i)},&
&\mathcal{O}_{hh\dots}^{(i)}.
\end{align}
Of course there are three classes of operators with one derivative:
\begin{align}
&\mathcal{O}_{f\dots}^{(i)},&
&\mathcal{O}_{g\dots}^{(i)},&
&\mathcal{O}_{h\dots}^{(i)}.
\end{align}

Referring to the restrictions of mixing under renormalization, all operators still obey the pattern displayed in Table \ref{tab:3}. One only has to keep in mind that a multiplet-index $D$ may actually represent an $f$, $g$ or $h$ as explained above. This completes our study of the symmetry properties for leading-twist three-quark operators with up to two derivatives. In the next section we give further identities for the special case of isospin $1/2$ symmetrized operators.

\section{Isospin Symmetrized Operators}
If one is interested in nucleon physics, one has to care about the appropriate isospin symmetrization of the operators used. Due to the presence of two equal quark flavors, identities will then show up which reduce the number of independent three-quark operators. We want to briefly discuss this in the following. Our results are summarized in Appendix \ref{App:5}.

There are two possible symmetry classes for isospin $1/2$: mixed antisymmetric, denoted by $MA$ in the following, and mixed symmetric ($MS$). Dealing with three quarks of isospin $I=1/2$, $m_I= \pm 1/2$, one can first couple two of them to either $m_I=0$ or $m_I=1$. For the proton the third quark field is then added in such a way that the resultant three-quark operator has $I=1/2$, $m_I=+1/2$:
\begin{eqnarray}
\vert MS \rangle & = & -\sqrt{\frac{2}{3}} \vert uud \rangle + \sqrt{\frac{1}{6}} (\left \vert udu \rangle +  \vert duu \rangle \right), \nonumber\\
\vert MA \rangle & = & \sqrt{\frac{1}{2}} \left( \vert udu \rangle - \vert duu \rangle \right).
\end{eqnarray}
The irreducible operators discussed in the previous section and listed in Appendix \ref{App:1} are converted to isospin $1/2$ operators when replacing the $f$, $g$ and $h$ quark fields by the appropriate $MS$ or $MA$ linear combinations of $uud$, $udu$ and $duu$ given above. The spinor and vector indices as well as the positions of the covariant derivatives remain unchanged. Let us give an example for the case of an $MA$ symmetrization:
\begin{align}
\mathcal{O}_{fg17}^{(4),MA} = \frac{5i}{8\sqrt{3}} &\left( \frac{3}{5} {(D\sigma)^{\{0}}_{\{ \dot 0} u^{0} {(D\sigma)^0}_{\dot 0 \}} d^{0}  u^{0 \}} \right. \nonumber\\ 
			  & - \frac{3}{5} {(D\sigma)^{\{0}}_{\{ \dot 0} d^{0} {(D\sigma)^0}_{\dot 0 \}} u^{0}  u^{0 \}}  \nonumber\\ 
                          & - {(D\sigma)^{\{1}}_{\{ \dot 0} u^{1} {(D\sigma)^1}_{\dot 0 \}} d^{1}  u^{0 \}} \nonumber\\
                          & + {(D\sigma)^{\{1}}_{\{ \dot 0} d^{1} {(D\sigma)^1}_{\dot 0 \}} u^{1}  u^{0 \}} \nonumber\\
                          & - 2 \cdot {(D\sigma)^{\{0} }_{\{ \dot 1} u^{1} {(D\sigma)^0}_{\dot 1 \}} d^{1}   u^{0 \}}  \nonumber\\
			  & \left. + 2 \cdot {(D\sigma)^{\{0} }_{\{ \dot 1} d^{1} {(D\sigma)^0}_{\dot 1 \}} u^{1}   u^{0 \}} \right).
\end{align}

The presence of two $u$ quarks leads to operator identities when use is made of the anticommutation relation for Grassmann variables and the symmetry of the coefficient tensors $T^{(i)}$. Again we want to clarify the procedure by a simple example. The $MA$ three-quark operators without derivative read after color antisymmetrization:
\begin{align}
\mathcal{O}^{(i),MA} &= T^{(i)}_{\alpha \beta \gamma} \frac{1}{\sqrt{2}} (u^a_\alpha d^b_\beta u^c_\gamma - d^a_\alpha  u^b_\beta u^c_\gamma) \epsilon_{abc} \nonumber\\
 &= \frac{1}{\sqrt{2}} (T^{(i)}_{\alpha \beta \gamma} - T^{(i)}_{\beta \alpha \gamma}) \cdot u^a_\alpha d^b_\beta u^c_\gamma \cdot \epsilon_{abc}.
 \label{MAid}
\end{align}
In Appendix \ref{App:2} we find that for the operators $\mathcal{O}^{(i)}_7$ and $\mathcal{O}^{(i)}_8$ the role of the spinor indices on the first and second quark is exchanged, i.e.,
\begin{align}
T^{(i)}_{7,\alpha \beta \gamma} = T^{(i)}_{8,\beta \alpha \gamma}.
\end{align}
When using this relation in (\ref{MAid}) the following identity between the two isospin symmetrized multiplets is derived:
\begin{align}
\mathcal{O}_7^{(i),MA} = - \mathcal{O}_8^{(i),MA}.
\end{align}

\begin{table}[t]
 \caption{Irreducibly transforming multiplets of three-quark operators with isospin $1/2$ sorted by their mass dimension.} 
 \renewcommand{\arraystretch}{1.7}
 \label{tab:5}
 \begin{tabular}{|c||c|c|c|}
   \hline
 & dimension 9/2 & dimension 11/2 & dimension 13/2\\
 & (0 derivatives) & (1 derivative) & (2 derivatives) \\
  \hline
  \hline
  $\tau_1^{\underline{4}}$ & $\begin{matrix} \mathcal{O}_{1}^{(i),MA},\\ \mathcal{O}_{3}^{(i),MA} \end{matrix}$ & & $\begin{matrix} \mathcal{O}^{(i),MA}_{ff1}, \\ \mathcal{O}^{(i),MA}_{ff2}, \, \mathcal{O}^{(i),MA}_{ff3} \end{matrix}$\\
  \hline
  $\tau_2^{\underline{4}}$ &  &  & $\begin{matrix} \mathcal{O}^{(i),MA}_{ff4}, \mathcal{O}^{(i),MA}_{ff5}, \\ \mathcal{O}_{ff6}^{(i),MA}, \, \mathcal{O}^{(i),MA}_{gh4}, \\ \mathcal{O}^{(i),MA}_{gh5}, \, \mathcal{O}_{gh6}^{(i),MA}\end{matrix}$\\
  \hline
  $\tau^{\underline{8}}$ & &  $\mathcal{O}^{(i),MA}_{f1}$ & $\begin{matrix} \mathcal{O}^{(i),MA}_{ff7},\, \mathcal{O}_{ff8}^{(i),MA}, \\ \mathcal{O}^{(i),MA}_{ff9}, \, \mathcal{O}^{(i),MA}_{gh7},\\ \mathcal{O}_{gh8}^{(i),MA}, \, \mathcal{O}^{(i),MA}_{gh9} \end{matrix}$ \\
  \hline
  $\tau_1^{\underline{12}}$ & $\mathcal{O}_{7}^{(i),MA}$ & $\begin{matrix} \mathcal{O}^{(i),MA}_{f2},\\ \mathcal{O}^{(i),MA}_{f3}, \\ \mathcal{O}^{(i),MA}_{f4} \end{matrix}$  & $\begin{matrix} \mathcal{O}^{(i),MA}_{ff10},\, \mathcal{O}^{(i),MA}_{ff11},\\ \mathcal{O}^{(i),MA}_{ff12},\, \mathcal{O}^{(i),MA}_{ff13}, \\ \mathcal{O}^{(i),MA}_{gh10},\, \mathcal{O}^{(i),MA}_{gh11},\\ \mathcal{O}^{(i),MA}_{gh12}, \, \mathcal{O}^{(i),MA}_{gh13} \end{matrix}$ \\
  \hline
  $\tau_2^{\underline{12}}$ &  &  $\begin{matrix} \mathcal{O}^{(i),MA}_{f5}, \\ \mathcal{O}^{(i),MA}_{f6},\\ \mathcal{O}^{(i),MA}_{f7}, \\ \mathcal{O}^{(i),MA}_{f8} \end{matrix}$ & $\begin{matrix} \mathcal{O}^{(i),MA}_{ff14}, \, \mathcal{O}^{(i),MA}_{ff15}, \\ \mathcal{O}^{(i),MA}_{ff16}, \, \mathcal{O}^{(i),MA}_{ff17}, \\ \mathcal{O}^{(i),MA}_{gh14}, \, \mathcal{O}^{(i),MA}_{gh15}, \\ \mathcal{O}^{(i),MA}_{gh16}, \, \mathcal{O}^{(i),MA}_{gh17} \end{matrix}$\\
  \hline
\end{tabular}
 \renewcommand{\arraystretch}{1.0}
\end{table}

A list of all identities induced by the isospin symmetrization is given in Appendix \ref{App:5}. There we systematically express all $MS$ operators in terms of $MA$ operators and then give all identities among the $MA$ operators. That leads to a minimal set of linearly independent three-quark operator multiplets with isospin $1/2$. These multiplets are summarized in Table \ref{tab:5}, where, just as in Table \ref{tab:3}, the allowed mixings under renormalization can be read off.

As all operators within one multiplet share the same renormalization coefficients, only an order of ten different renormalization matrices of dimension eight by eight or lower need to be calculated. The nonperturbative evaluation of these coefficients for two flavors of clover fermions is in progress and will yield a full set of
renormalization constants for isospin 1/2 symmetrized operators of leading twist with up to two derivatives. These upcoming results will allow us to renormalize the first few moments of the proton distribution amplitude. To this end, one relates matrix elements of identically transforming three-quark operators, e.g., $\mathcal{O}_1 = \mathcal{O}^{(i),MA}_{f5}$, $\mathcal{O}_2 = \mathcal{O}^{(i),MA}_{f6}$ and $\mathcal{O}_3 = \mathcal{O}^{(i),MA}_{f7}$ with $i$ fixed, to moments $\Phi_1$, $\Phi_2$ and $\Phi_3$ of the proton distribution amplitude.
Schematically we can express the renormalized moments in terms of the renormalized operators by 
\begin{align}
\Phi_1^{\mathrm {ren}} \sim & \,
          \langle 0 | \mathcal{O}_1^{\mathrm {ren}}  | P \rangle \,, \nonumber \\
\Phi_2^{\mathrm {ren}} \sim & \,
          \langle 0 | \mathcal{O}_2^{\mathrm {ren}}  | P \rangle \,, \nonumber \\
\Phi_3^{\mathrm {ren}} \sim & \,
          \langle 0 | \mathcal{O}_3^{\mathrm {ren}}  | P \rangle \,, \nonumber 
\end{align}
where $| P \rangle$ denotes a proton state of definite momentum (for details see \cite{Niko2008}).
With the renormalization and mixing coefficients of our three-quark operators, $\mathcal{O}_i^{\mathrm {ren}} = Z_{ij} \mathcal{O}_j$, we finally arrive at the following relation between the renormalized and the bare moments: 
\begin{align}
\Phi_1^{\mathrm {ren}}  = & \,
        Z_{11} \Phi_1 + Z_{12} \Phi_2 + Z_{13} \Phi_3 \,, \nonumber \\
\Phi_2^{\mathrm {ren}}  = & \,
        Z_{21} \Phi_1 + Z_{22} \Phi_2 + Z_{23} \Phi_3 \,, \nonumber \\
\Phi_3^{\mathrm {ren}}  = & \,
        Z_{31} \Phi_1 + Z_{32} \Phi_2 + Z_{33} \Phi_3 \,. \nonumber 
\end{align}
This emphasizes once more the importance of the detailed study of operator mixing under renormalization on the lattice that we have presented here.

\section{Summary and Outlook}
\label{sec:8}
Three-quark operators play an important role in the lattice determination of non-perturbative contributions to hard exclusive processes involving baryons. In order to get continuum results from calculations using these operators, they must be renormalized and it is crucial to study the mixing behavior.

\begin{sloppypar}
Here we have investigated the constraints imposed by group theory. The hypercubic group and its spinorial analogue were used to determine the behavior of three-quark operators under transformations of the discretized space-time. In a first step we substantially reduced the dimensions of the occurring representation matrices by studying the continuum behavior. Then we used a projector to derive the multiplets of $\overline{H(4)}$ irreducibly transforming three-quark operators.
When grouping them according to their representation and mass dimension, the possible mixing patterns can be read off. This provides the basis for the renormalization of iso\-spin-symmetrized three-quark operators needed for the evaluation of nucleon distribution amplitudes, which will be published in a forthcoming paper.
\end{sloppypar}

\appendix
\section{Irreducibly Transforming Three-Quark Operators}
\label{App:1}
In this Appendix we list the explicit form of the $\overline{H(4)}$ irreducibly transforming multiplets of three-quark operators. The operators are constructed such that under group action of $\overline{H(4)}$ any $i$-th operator within one multiplet transforms identically to the $i$-th operator of another multiplet belonging to an equivalent representation.

On demand, interested groups may also receive the coefficient tensors in electronic form to facilitate the implementation of the corresponding operators.

\subsection{Operators without Derivatives}
\label{App:2}
For three-quark operators without derivatives we present the full set of irreducible operators, including those of non-leading twist $\left( \mathcal{O}_1^{(i)} \right.$, $\mathcal{O}_2^{(i)}$, $\mathcal{O}_3^{(i)}$, $\mathcal{O}_4^{(i)}$ and $\left. \mathcal{O}_5^{(i)} \right)$. The spinor indices are given in the chiral Weyl representation.

The first five multiplets belong to the irreducible representation $\tau^{\underline{4}}_1$:
\begin{align}
\mathcal{O}_1^{(1)} &= \frac{1}{\sqrt{6}} (f^1g^0h^0 - 2\cdot f^0g^1h^0 + f^0g^0h^1), \nonumber\\
\mathcal{O}_1^{(2)} &= \frac{1}{\sqrt{6}} (2\cdot f^1g^0h^1 - f^0g^1h^1 - f^1g^1h^0), \nonumber\\
\mathcal{O}_1^{(3)} &= \frac{1}{\sqrt{6}} (f_{\dot 1}g_{\dot 0}h_{\dot 0} - 2\cdot f_{\dot 0}g_{\dot 1}h_{\dot 0} + f_{\dot 0}g_{\dot 0}h_{\dot 1}), \nonumber\\
\mathcal{O}_1^{(4)} &= \frac{1}{\sqrt{6}} (2\cdot f_{\dot 1}g_{\dot 0}h_{\dot 1} - f_{\dot 0}g_{\dot 1}h_{\dot 1} - f_{\dot 1}g_{\dot 1}h_{\dot 0}),
\end{align}
\begin{align}
\mathcal{O}_2^{(1)} &= \frac{1}{\sqrt{6}} (f^1g^0h^0 + f^0g^1h^0 - 2\cdot f^0g^0h^1), \nonumber\\
\mathcal{O}_2^{(2)} &= \frac{1}{\sqrt{6}} (2\cdot f^1g^1h^0 - f^0g^1h^1 - f^1g^0h^1), \nonumber\\
\mathcal{O}_2^{(3)} &= \frac{1}{\sqrt{6}} (f_{\dot 1}g_{\dot 0}h_{\dot 0} + f_{\dot 0}g_{\dot 1}h_{\dot 0} - 2\cdot f_{\dot 0}g_{\dot 0}h_{\dot 1}), \nonumber\\
\mathcal{O}_2^{(4)} &= \frac{1}{\sqrt{6}} (2\cdot f_{\dot 1}g_{\dot 1}h_{\dot 0} - f_{\dot 0}g_{\dot 1}h_{\dot 1} - f_{\dot 1}g_{\dot 0}h_{\dot 1}),
\end{align}
\begin{align}
\mathcal{O}_3^{(1)} &= \frac{1}{\sqrt{2}} (f^0g_{\dot 0}h_{\dot 1} - f^0g_{\dot 1}h_{\dot 0}), \nonumber\\
\mathcal{O}_3^{(2)} &= \frac{1}{\sqrt{2}} (f^1g_{\dot 0}h_{\dot 1} - f^1g_{\dot 1}h_{\dot 0}), \nonumber\\ 
\mathcal{O}_3^{(3)} &= \frac{1}{\sqrt{2}} (f_{\dot 0}g^0h^1 - f_{\dot 0}g^1h^0), \nonumber\\
\mathcal{O}_3^{(4)} &= \frac{1}{\sqrt{2}} (f_{\dot 1}g^0h^1 - f_{\dot 1}g^1h^0).
\end{align}
The operators $\mathcal{O}_4^{(i)}$ ($\mathcal{O}_5^{(i)}$) are constructed from $\mathcal{O}_3^{(i)}$ by interchange of the indices on the quarks $f$ and $g$ ($h$). They are chirality partners: whereas $\mathcal{O}_3^{(i)}$ originates from a $-++$ chirality combination of the three quark fields, $\mathcal{O}_4^{(i)}$ and $\mathcal{O}_5^{(i)}$ come from $+-+$ and $++-$ combinations, respectively (cf. (\ref{SO4_chiral})). 

We proceed with the leading-twist operators. The operators $\mathcal{O}_6^{(i)}$ contain three quark fields of equal chirality and transform according to $\tau^{\underline{8}}$:
\begin{align}
\mathcal{O}_6^{(1)} &= f^0g^0h^0, \nonumber\\
\mathcal{O}_6^{(2)} &= \frac{1}{\sqrt{3}} (f^0g^0h^1 + f^0g^1h^0 + f^1g^0h^0), \nonumber\\ 
\mathcal{O}_6^{(3)} &= \frac{1}{\sqrt{3}} (f^0g^1h^1 + f^1g^0h^1 + f^1g^1h^0), \nonumber\\
\mathcal{O}_6^{(4)} &= f^1g^1h^1, \nonumber\\
\mathcal{O}_6^{(5)} &= f_{\dot 0}g_{\dot 0}h_{\dot 0}, \nonumber\\
\mathcal{O}_6^{(6)} &= \frac{1}{\sqrt{3}} (f_{\dot 0}g_{\dot 0}h_{\dot 1} + f_{\dot 0}g_{\dot 1}h_{\dot 0} + f_{\dot 1}g_{\dot 0}h_{\dot 0}), \nonumber\\
\mathcal{O}_6^{(7)} &= \frac{1}{\sqrt{3}} (f_{\dot 0}g_{\dot 1}h_{\dot 1} + f_{\dot 1}g_{\dot 0}h_{\dot 1} + f_{\dot 1}g_{\dot 1}h_{\dot 0}), \nonumber\\
\mathcal{O}_6^{(8)} &= f_{\dot 1}g_{\dot 1}h_{\dot 1}.
\end{align}
Finally, there are three more multiplets that belong to $\tau^{\underline{12}}_1$:
\begin{align}
\mathcal{O}_7^{(1)} &= f^0g_{\dot 0}h_{\dot 0}, \nonumber\\
\mathcal{O}_7^{(2)} &= \frac{1}{\sqrt{2}} (f^0g_{\dot 0}h_{\dot 1} + f^0g_{\dot 1}h_{\dot 0}), \nonumber\\
\mathcal{O}_7^{(3)} &= f^0g_{\dot 1}h_{\dot 1}, \nonumber\\
\mathcal{O}_7^{(4)} &= f^1g_{\dot 0}h_{\dot 0}, \nonumber\\
\mathcal{O}_7^{(5)} &= \frac{1}{\sqrt{2}} (f^1g_{\dot 0}h_{\dot 1}+f^1g_{\dot 1}h_{\dot 0}), \nonumber\\
\mathcal{O}_7^{(6)} &= f^1g_{\dot 1}h_{\dot 1}, \nonumber\\
\mathcal{O}_7^{(7)} &= f_{\dot 0}g^0h^0, \nonumber\\
\mathcal{O}_7^{(8)} &= \frac{1}{\sqrt{2}} (f_{\dot 0}g^0h^1+f_{\dot 0}g^1h^0), \nonumber\\
\mathcal{O}_7^{(9)} &= f_{\dot 0}g^1h^1, \nonumber\\
\mathcal{O}_7^{(10)} &= f_{\dot 1}g^0h^0, \nonumber\\
\mathcal{O}_7^{(11)} &= \frac{1}{\sqrt{2}} (f_{\dot 1}g^0h^1+f_{\dot 1}g^1h^0), \nonumber\\
\mathcal{O}_7^{(12)} &= f_{\dot 1}g^1h^1.
\end{align}
\begin{sloppypar}
The operators $\mathcal{O}_8^{(i)}$ ($\mathcal{O}_9^{(i)}$) are chirality partners of $\mathcal{O}_7^{(i)}$. They follow when exchanging the index on quarks one and two (three).
\end{sloppypar}

\subsection{Operators with one Derivative}
\label{App:3}
We list the operators with one and two derivatives using the dotted and undotted indices for the quark fields as introduced in Sec. \ref{sec:5} and denote separate total symmetrization in the (un)dotted indices by curly brackets. The product of the covariant derivatives with the Pauli matrices reads in the Euclidean formulation
\begin{align}
{(D \sigma)^0}_{\dot 0} = +D_1-i D_2, \nonumber\\
{(D \sigma)^0}_{\dot 1} = -D_3+i D_4, \nonumber\\
{(D \sigma)^1}_{\dot 0} = -D_3-i D_4, \nonumber\\
{(D \sigma)^1}_{\dot 1} = -D_1-i D_2.
\end{align}

The eight operators $\mathcal{O}_{D1}^{(i)}$ belong to the irreducible representation $\tau^{\underline{8}}$ and are constructed from three quarks with equal chiralities:
\footnotesize
\begin{align}
\mathcal{O}_{D1}^{(1)} &= +\frac{1}{2} \cdot \left( f_{\{ \dot 0} g_{\dot 0} {(D\sigma)^1}_{\dot 0} h_{\dot 0\}} + f_{\{ \dot 1} g_{\dot 1} {(D\sigma)^1}_{\dot 1} h_{\dot 1\}} \right), \nonumber\\
\mathcal{O}_{D1}^{(2)} &= -\sqrt{3} \cdot f_{\{ \dot 1} g_{\dot 1} {(D\sigma)^0}_{\dot 0} h_{\dot 0\}}, \nonumber\\
\mathcal{O}_{D1}^{(3)} &= +\sqrt{3} \cdot f_{\{ \dot 1} g_{\dot 1} {(D\sigma)^1}_{\dot 0} h_{\dot 0\}}, \nonumber\\
\mathcal{O}_{D1}^{(4)} &= - \frac{1}{2} \cdot \left( f_{\{ \dot 0} g_{\dot 0} {(D\sigma)^0}_{\dot 0} h_{\dot 0\}} +  f_{\{ \dot 1} g_{\dot 1} {(D\sigma)^0}_{\dot 1} h_{\dot 1\}} \right), \nonumber\\
\mathcal{O}_{D1}^{(5)} &= - \frac{1}{2} \cdot \left( f^{\{ 0} g^{0} {(D\sigma)^0}_{\dot 1} h^{0\}} + f^{\{ 1} g^{1} {(D\sigma)^1}_{\dot 1} h^{1\}} \right), \nonumber\\
\mathcal{O}_{D1}^{(6)} &= +\sqrt{3} \cdot f^{\{ 1} g^{1} {(D\sigma)^0}_{\dot 0} h^{0\}}, \nonumber\\
\mathcal{O}_{D1}^{(7)} &= -\sqrt{3} \cdot f^{\{ 1} g^{1} {(D\sigma)^0}_{\dot 1} h^{0\}}, \nonumber\\
\mathcal{O}_{D1}^{(8)} &= +\frac{1}{2} \cdot \left( f^{\{ 0} g^{0} {(D\sigma)^0}_{\dot 0} h^{0\}} + f^{\{ 1} g^{1} {(D\sigma)^1}_{\dot 0} h^{1\}} \right).
\end{align}
\normalsize

The twelve operators $\mathcal{O}_{D2}^{(i)}$ generate a $\tau_1^{\underline{12}}$ irreducible representation and arise from quark chiralities $-++$:
\footnotesize
\begin{align}
\mathcal{O}_{D2}^{(1)} &= -\frac{\sqrt{3}}{2\sqrt{2}} \cdot (f_{\{\dot 0} g^{\{1} {(D\sigma)^0}_{\dot 0\}} h^{0\}}  + f_{\{\dot 1} g^{\{1} {(D\sigma)^1}_{\dot 1\}} h^{1\}} ),\nonumber\\
\mathcal{O}_{D2}^{(2)} &= \sqrt{3} \cdot f_{\{\dot 1} g^{\{1} {(D\sigma)^0}_{\dot 0\}} h^{0\}}, \nonumber\\
\mathcal{O}_{D2}^{(3)} &= -\frac{\sqrt{3}}{2\sqrt{2}} \cdot (f_{\{\dot 0} g^{\{1} {(D\sigma)^1}_{\dot 0\}} h^{1\}} +  f_{\{\dot 1} g^{\{1} {(D\sigma)^0}_{\dot 1\}} h^{0\}} ), \nonumber\\
\mathcal{O}_{D2}^{(4)} &= \frac{\sqrt{3}}{2\sqrt{2}} \cdot ( f_{\{\dot 0} g^{\{1} {(D\sigma)^1}_{\dot 0\}} h^{0\}} +  f_{\{\dot 1} g^{\{0} {(D\sigma)^0}_{\dot 1\}} h^{0\}} ), \nonumber\\
\mathcal{O}_{D2}^{(5)} &= -\sqrt{3} f_{\{\dot 1} g^{\{1} {(D\sigma)^1}_{\dot 0\}} h^{0\}}, \nonumber\\
\mathcal{O}_{D2}^{(6)} &= \frac{\sqrt{3}}{2\sqrt{2}} \cdot ( f_{\{\dot 0} g^{\{0} {(D\sigma)^0}_{\dot 0\}} h^{0\}} +  f_{\{\dot 1} g^{\{1} {(D\sigma)^1}_{\dot 1\}} h^{0\}} ), \nonumber\\
\mathcal{O}_{D2}^{(7)} &= \frac{\sqrt{3}}{2\sqrt{2}} \cdot ( f^{\{0} g_{\{\dot 1} {(D\sigma)^{0\}}}_{\dot 0} h_{\dot 0\}}  + f^{\{1} g_{\{\dot 1} {(D\sigma)^{1\}}}_{\dot 1} h_{\dot 1\}} ), \nonumber\\
\mathcal{O}_{D2}^{(8)} &= -\sqrt{3} \cdot f^{\{1} g_{\{\dot 1} {(D\sigma)^{0\}}}_{\dot 0} h_{\dot 0\}}, \nonumber\\
\mathcal{O}_{D2}^{(9)} &= \frac{\sqrt{3}}{2\sqrt{2}} \cdot ( f^{\{0} g_{\{\dot 1} {(D\sigma)^{0\}}}_{\dot 1} h_{\dot 1\}} + f^{\{1} g_{\{\dot 1} {(D\sigma)^{1\}}}_{\dot 0} h_{\dot 0\}} ), \nonumber\\
\mathcal{O}_{D2}^{(10)} &= -\frac{\sqrt{3}}{2\sqrt{2}} \cdot ( f^{\{0} g_{\{\dot 1} {(D\sigma)^{0\}}}_{\dot 1} h_{\dot 0\}} + f^{\{1} g_{\{\dot 0} {(D\sigma)^{1\}}}_{\dot 0} h_{\dot 0\}} ), \nonumber\\
\mathcal{O}_{D2}^{(11)} &= \sqrt{3} \cdot f^{\{1} g_{\{\dot 1} {(D\sigma)^{0\}}}_{\dot 1} h_{\dot 0\}}, \nonumber\\
\mathcal{O}_{D2}^{(12)} &= -\frac{\sqrt{3}}{2\sqrt{2}} \cdot ( f^{\{0} g_{\{\dot 0} {(D\sigma)^{0\}}}_{\dot 0} h_{\dot 0\}} + f^{\{1} g_{\{\dot 1} {(D\sigma)^{1\}}}_{\dot 1} h_{\dot 0\}}).
\end{align}
\normalsize
The chirality partners $\mathcal{O}_{D3}^{(i)}$ ($\mathcal{O}_{D4}^{(i)}$) are derived from $\mathcal{O}_{D2}^{(i)}$ by exchanging the indices assigned to the $f$ quark with those of $g$ ($h$).

The last four multiplets of operators transform according to the irreducible representation $\tau_2^{\underline{12}}$. The operators $\mathcal{O}_{D5}^{(i)}$ are constructed from quark chiralities $-++$:
\footnotesize
\begin{align}
\mathcal{O}_{D5}^{(1)} &= \frac{1}{2\sqrt{2}} f_{\{\dot 0} g^{\{0} {(D\sigma)^0}_{\dot 0\}} h^{0\}} - \frac{3}{2\sqrt{2}} f_{\{\dot 1} g^{\{1} {(D\sigma)^1}_{\dot 1\}} h^{0\}}, \nonumber\\
\mathcal{O}_{D5}^{(2)} &= \frac{3}{2\sqrt{2}} f_{\{\dot 0} g^{\{1} {(D\sigma)^0}_{\dot 0\}} h^{0\}} - \frac{1}{2\sqrt{2}} f_{\{\dot 1} g^{\{1} {(D\sigma)^1}_{\dot 1\}} h^{1\}}, \nonumber\\
\mathcal{O}_{D5}^{(3)} &= \frac{3}{2\sqrt{2}} f_{\{\dot 0} g^{\{1} {(D\sigma)^1}_{\dot 0\}} h^{0\}} - \frac{1}{2\sqrt{2}} f_{\{\dot 1} g^{\{0} {(D\sigma)^0}_{\dot 1\}} h^{0\}}, \nonumber\\
\mathcal{O}_{D5}^{(4)} &= \frac{1}{2\sqrt{2}} f_{\{\dot 0} g^{\{1} {(D\sigma)^1}_{\dot 0\}} h^{1\}} - \frac{3}{2\sqrt{2}} f_{\{\dot 1} g^{\{1} {(D\sigma)^0}_{\dot 1\}} h^{0\}}, \nonumber\\
\mathcal{O}_{D5}^{(5)} &= f_{\{\dot 1} g^{\{0} {(D\sigma)^0}_{\dot 0\}} h^{0\}}, \nonumber\\
\mathcal{O}_{D5}^{(6)} &= f_{\{\dot 1} g^{\{1} {(D\sigma)^1}_{\dot 0\}} h^{1\}}, \nonumber\\
\mathcal{O}_{D5}^{(7)} &= \frac{1}{2\sqrt{2}} f^{\{0} g_{\{\dot 0} {(D\sigma)^{0\}}}_{\dot 0} h_{\dot 0\}} - \frac{3}{2\sqrt{2}} f^{\{1} g_{\{\dot 1} {(D\sigma)^{1\}}}_{\dot 1} h_{\dot 0\}}, \nonumber\\
\mathcal{O}_{D5}^{(8)} &= \frac{3}{2\sqrt{2}} f^{\{0} g_{\{\dot 1} {(D\sigma)^{0\}}}_{\dot 0} h_{\dot 0\}} - \frac{1}{2\sqrt{2}} f^{\{1} g_{\{\dot 1} {(D\sigma)^{1\}}}_{\dot 1} h_{\dot 1\}}, \nonumber\\
\mathcal{O}_{D5}^{(9)} &= \frac{3}{2\sqrt{2}} f^{\{0} g_{\{\dot 1} {(D\sigma)^{0\}}}_{\dot 1} h_{\dot 0\}} - \frac{1}{2\sqrt{2}} f^{\{1} g_{\{\dot 0} {(D\sigma)^{1\}}}_{\dot 0} h_{\dot 0\}}, \nonumber\\
\mathcal{O}_{D5}^{(10)} &= \frac{1}{2\sqrt{2}} f^{\{0} g_{\{\dot 1} {(D\sigma)^{0\}}}_{\dot 1} h_{\dot 1\}} - \frac{3}{2\sqrt{2}} f^{\{1} g_{\{\dot 1} {(D\sigma)^{1\}}}_{\dot 0} h_{\dot 0\}}, \nonumber\\
\mathcal{O}_{D5}^{(11)} &= f^{\{1} g_{\{\dot 0} {(D\sigma)^{0\}}}_{\dot 0} h_{\dot 0\}}, \nonumber\\
\mathcal{O}_{D5}^{(12)} &= f^{\{1} g_{\{\dot 1} {(D\sigma)^{0\}}}_{\dot 1} h_{\dot 1\}}.
\end{align}
\normalsize
$\mathcal{O}_{D6}^{(i)}$ ($\mathcal{O}_{D7}^{(i)}$) result from $\mathcal{O}_{D5}^{(i)}$ by interchanging the indices on the first and second (third) quark. Finally we have:
\footnotesize
\begin{align}
\mathcal{O}_{D8}^{(1)} &= +\sqrt{2} f_{\{ \dot 1} g_{\dot 1} {(D\sigma)^1}_{\dot 1} h_{\dot 0\}}, \nonumber\\
\mathcal{O}_{D8}^{(2)} &= -\sqrt{2} f_{\{ \dot 1} g_{\dot 0} {(D\sigma)^0}_{\dot 0} h_{\dot 0\}}, \nonumber\\
\mathcal{O}_{D8}^{(3)} &= +\sqrt{2} f_{\{ \dot 1} g_{\dot 0} {(D\sigma)^1}_{\dot 0} h_{\dot 0\}}, \nonumber\\
\mathcal{O}_{D8}^{(4)} &= -\sqrt{2} f_{\{ \dot 1} g_{\dot 1} {(D\sigma)^0}_{\dot 1} h_{\dot 0\}}, \nonumber\\
\mathcal{O}_{D8}^{(5)} &= +\frac{1}{2} f_{\{ \dot 1} g_{\dot 1} {(D\sigma)^1}_{\dot 1} h_{\dot 1\}} - \frac{1}{2} f_{\{ \dot 0} g_{\dot 0} {(D\sigma)^1}_{\dot 0} h_{\dot 0\}}, \nonumber\\
\mathcal{O}_{D8}^{(6)} &= +\frac{1}{2} f_{\{ \dot 0} g_{\dot 0} {(D\sigma)^0}_{\dot 0} h_{\dot 0\}} - \frac{1}{2} f_{\{ \dot 1} g_{\dot 1} {(D\sigma)^0}_{\dot 1} h_{\dot 1\}}, \nonumber\\
\mathcal{O}_{D8}^{(7)} &= +\sqrt{2} f^{\{ 1} g^{1} {(D\sigma)^1}_{\dot 1} h^{0\}}, \nonumber\\
\mathcal{O}_{D8}^{(8)} &= -\sqrt{2} f^{\{ 1} g^{0} {(D\sigma)^0}_{\dot 0} h^{0\}}, \nonumber\\
\mathcal{O}_{D8}^{(9)} &= +\sqrt{2} f^{\{ 1} g^{0} {(D\sigma)^0}_{\dot 1} h^{0\}}, \nonumber\\
\mathcal{O}_{D8}^{(10)} &= -\sqrt{2} f^{\{ 1} g^{1} {(D\sigma)^1}_{\dot 0} h^{0\}}, \nonumber\\
\mathcal{O}_{D8}^{(11)} &= +\frac{1}{2} f^{\{ 1} g^{1} {(D\sigma)^1}_{\dot 1} h^{1\}} - \frac{1}{2} f^{\{ 0} g^{0} {(D\sigma)^0}_{\dot 1} h^{0\}}, \nonumber\\
\mathcal{O}_{D8}^{(12)} &= +\frac{1}{2} f^{\{ 0} g^{0} {(D\sigma)^0}_{\dot 0} h^{0\}} - \frac{1}{2} f^{\{ 1} g^{1} {(D\sigma)^1}_{\dot 0} h^{1\}}.
\end{align}
\normalsize

\subsection{Operators with two Derivatives}
\label{App:4}
Here we display irreducible multiplets of three-quark operators with two covariant derivatives. As stated in the text, the positions of the derivatives do not influence the transformation properties. Hence one can produce further multiplets by assigning the derivatives to any quark field one likes. They can also act on two different quarks.

The first three multiplets transform according to $\tau^{\underline{4}}_1$:
\scriptsize
\begin{align}
\mathcal{O}_{DD1}^{(1)} =& +\frac{3}{2} f^{\{1}g_{\{ \dot 1} {(D\sigma)^0}_{\dot 1} {(D\sigma)^{0\}}}_{\dot 0} h_{\dot 0 \}} + \frac{1}{4} f^{\{1}g_{\{ \dot 0} {(D\sigma)^1}_{\dot 0} {(D\sigma)^{1\}}}_{\dot 0} h_{\dot 0 \}} \nonumber\\ & + \frac{1}{4} f^{\{1}g_{\{ \dot 1} {(D\sigma)^1}_{\dot 1} {(D\sigma)^{1\}}}_{\dot 1} h_{\dot 1 \}}, \nonumber\\
\mathcal{O}_{DD1}^{(2)} =& -\frac{1}{4} f^{\{0}g_{\{ \dot 0} {(D\sigma)^0}_{\dot 0} {(D\sigma)^{0\}}}_{\dot 0} h_{\dot 0 \}} -\frac{1}{4} f^{\{0}g_{\{ \dot 1} {(D\sigma)^0}_{\dot 1} {(D\sigma)^{0\}}}_{\dot 1} h_{\dot 1 \}} \nonumber\\ &  - \frac{3}{2} f^{\{1}g_{\{ \dot 1} {(D\sigma)^1}_{\dot 1} {(D\sigma)^{0\}}}_{\dot 0} h_{\dot 0 \}}, \nonumber\\
\mathcal{O}_{DD1}^{(3)} =& +\frac{3}{2} f_{\{\dot 1}g^{\{1} {(D\sigma)^1}_{\dot 0} {(D\sigma)^0}_{\dot 0 \}} h^{0\}} + \frac{1}{4} f_{\{\dot 1}g^{\{ 0} {(D\sigma)^0}_{\dot 1} {(D\sigma)^0}_{\dot 1 \}} h^{0\}} \nonumber\\ & + \frac{1}{4} f_{\{\dot 1}g^{\{ 1} {(D\sigma)^1}_{\dot 1} {(D\sigma)^1}_{\dot 1 \}} h^{1\}}, \nonumber\\
\mathcal{O}_{DD1}^{(4)} =& -\frac{1}{4} f_{\{\dot 0}g^{\{0} {(D\sigma)^0}_{\dot 0} {(D\sigma)^0}_{\dot 0 \}} h^{0\}} -\frac{1}{4} f_{\{\dot 0}g^{\{1} {(D\sigma)^1}_{\dot 0} {(D\sigma)^1}_{\dot 0 \}} h^{1\}} \nonumber\\ & - \frac{3}{2} f_{\{\dot 1}g^{\{1} {(D\sigma)^1}_{\dot 1} {(D\sigma)^0}_{\dot 0 \}} h^{0\}}.
\end{align}
\normalsize
The operators $\mathcal{O}_{DD2}^{(i)}$ ($\mathcal{O}_{DD3}^{(i)}$) are generated by exchanging the index on the first with that on the second (third) quark field.
\scriptsize
\begin{align}
\mathcal{O}_{DD4}^{(1)} =& \frac{\sqrt{3}}{2} f_{\{\dot 0}g^{\{1} {(D\sigma)^1}_{\dot 0} {(D\sigma)^0}_{\dot 0 \}} h^{0\}} - \frac{\sqrt{3}}{4} f_{\{\dot 1}g^{\{0} {(D\sigma)^0}_{\dot 1} {(D\sigma)^0}_{\dot 0 \}} h^{0\}} \nonumber\\ & -\frac{\sqrt{3}}{4} f_{\{\dot 1}g^{\{1} {(D\sigma)^1}_{\dot 1} {(D\sigma)^1}_{\dot 0 \}} h^{1\}}, \nonumber\\
\mathcal{O}_{DD4}^{(2)} =& \frac{\sqrt{3}}{4} f_{\{\dot 1}g^{\{0} {(D\sigma)^0}_{\dot 0} {(D\sigma)^0}_{\dot 0 \}} h^{0\}} + \frac{\sqrt{3}}{4} f_{\{\dot 1}g^{\{1} {(D\sigma)^1}_{\dot 0} {(D\sigma)^1}_{\dot 0 \}} h^{1\}} \nonumber\\ & -\frac{\sqrt{3}}{2} f_{\{\dot 1}g^{\{ 1} {(D\sigma)^1}_{\dot 1} {(D\sigma)^0}_{\dot 1 \}} h^{0\}}, \nonumber\\
\mathcal{O}_{DD4}^{(3)} =& \frac{\sqrt{3}}{2} f^{\{0}g_{\{ \dot 1} {(D\sigma)^0}_{\dot 1} {(D\sigma)^{0\}}}_{\dot 0} h_{\dot 0 \}} - \frac{\sqrt{3}}{4} f^{\{1}g_{\{ \dot 0} {(D\sigma)^1}_{\dot 0} {(D\sigma)^{0\}}}_{\dot 0} h_{\dot 0 \}} \nonumber\\ &  -\frac{\sqrt{3}}{4} f^{\{1}g_{\{ \dot 1} {(D\sigma)^1}_{\dot 1} {(D\sigma)^{0\}}}_{\dot 1} h_{\dot 1 \}}, \nonumber\\
\mathcal{O}_{DD4}^{(4)} =& \frac{\sqrt{3}}{4} f^{\{1}g_{\{ \dot 0} {(D\sigma)^0}_{\dot 0} {(D\sigma)^{0\}}}_{\dot 0} h_{\dot 0 \}} + \frac{\sqrt{3}}{4} f^{\{1}g_{\{ \dot 1} {(D\sigma)^0}_{\dot 1} {(D\sigma)^{0\}}}_{\dot 1} h_{\dot 1 \}} \nonumber\\ & -\frac{\sqrt{3}}{2} f^{\{1}g_{\{ \dot 1} {(D\sigma)^1}_{\dot 1} {(D\sigma)^{1\}}}_{\dot 0} h_{\dot 0 \}}.
\end{align}
\normalsize
Again, $\mathcal{O}_{DD5}^{(i)}$ ($\mathcal{O}_{DD6}^{(i)}$) result from $\mathcal{O}_{DD4}^{(i)}$ by exchanging the index on the $f$ with that on the $g$ ($h$) quark field. They belong to $\tau^{\underline{4}}_2$, whereas the following three multiplets transform according to $\tau^{\underline{8}}$:
\scriptsize
\begin{align}
\mathcal{O}_{DD7}^{(1)} =& \frac{1}{4} f_{\{\dot 0}g^{\{0} {(D\sigma)^0}_{\dot 0} {(D\sigma)^0}_{\dot 0 \}} h^{0\}} + \frac{1}{4} f_{\{\dot 0}g^{\{1} {(D\sigma)^1}_{\dot 0} {(D\sigma)^1}_{\dot 0 \}} h^{1\}} \nonumber\\ & - \frac{3}{2} f_{\{\dot 1}g^{\{1} {(D\sigma)^1}_{\dot 1} {(D\sigma)^0}_{\dot 0 \}} h^{0\}}, \nonumber\\
\mathcal{O}_{DD7}^{(2)} =& \frac{\sqrt{3}}{2} f_{\{\dot 0}g^{\{1} {(D\sigma)^1}_{\dot 0} {(D\sigma)^0}_{\dot 0 \}} h^{0\}} + \frac{\sqrt{3}}{4} f_{\{\dot 1}g^{\{0} {(D\sigma)^0}_{\dot 1} {(D\sigma)^0}_{\dot 0 \}} h^{0\}} \nonumber\\ &+  \frac{\sqrt{3}}{4} f_{\{\dot 1}g^{\{1} {(D\sigma)^1}_{\dot 1} {(D\sigma)^1}_{\dot 0 \}} h^{1\}}, \nonumber\\
\mathcal{O}_{DD7}^{(3)} =& \frac{\sqrt{3}}{4} f_{\{\dot 1}g^{\{0} {(D\sigma)^0}_{\dot 0} {(D\sigma)^0}_{\dot 0 \}} h^{0\}} + \frac{\sqrt{3}}{4} f_{\{\dot 1}g^{\{1} {(D\sigma)^1}_{\dot 0} {(D\sigma)^1}_{\dot 0 \}} h^{1\}} \nonumber\\ &+  \frac{\sqrt{3}}{2} f_{\{\dot 1}g^{\{ 1} {(D\sigma)^1}_{\dot 1} {(D\sigma)^0}_{\dot 1 \}} h^{0\}}, \nonumber\\
\mathcal{O}_{DD7}^{(4)} =& \frac{3}{2} f_{\{\dot 1}g^{\{1} {(D\sigma)^1}_{\dot 0} {(D\sigma)^0}_{\dot 0 \}} h^{0\}} - \frac{1}{4} f_{\{\dot 1}g^{\{ 0} {(D\sigma)^0}_{\dot 1} {(D\sigma)^0}_{\dot 1 \}} h^{0\}} \nonumber\\ & - \frac{1}{4} f_{\{\dot 1}g^{\{ 1} {(D\sigma)^1}_{\dot 1} {(D\sigma)^1}_{\dot 1 \}} h^{1\}}, \nonumber\\
\mathcal{O}_{DD7}^{(5)} =& \frac{1}{4} f^{\{0}g_{\{ \dot 0} {(D\sigma)^0}_{\dot 0} {(D\sigma)^{0\}}}_{\dot 0} h_{\dot 0 \}} + \frac{1}{4} f^{\{0}g_{\{ \dot 1} {(D\sigma)^0}_{\dot 1} {(D\sigma)^{0\}}}_{\dot 1} h_{\dot 1 \}} \nonumber\\ & - \frac{3}{2} f^{\{1}g_{\{ \dot 1} {(D\sigma)^1}_{\dot 1} {(D\sigma)^{0\}}}_{\dot 0} h_{\dot 0 \}}, \nonumber\\
\mathcal{O}_{DD7}^{(6)} =& \frac{\sqrt{3}}{2} f^{\{0}g_{\{ \dot 1} {(D\sigma)^0}_{\dot 1} {(D\sigma)^{0\}}}_{\dot 0} h_{\dot 0 \}} + \frac{\sqrt{3}}{4} f^{\{1}g_{\{ \dot 0} {(D\sigma)^1}_{\dot 0} {(D\sigma)^{0\}}}_{\dot 0} h_{\dot 0 \}} \nonumber\\ & + \frac{\sqrt{3}}{4} f^{\{1}g_{\{ \dot 1} {(D\sigma)^1}_{\dot 1} {(D\sigma)^{0\}}}_{\dot 1} h_{\dot 1 \}}, \nonumber\\
\mathcal{O}_{DD7}^{(7)} =& \frac{\sqrt{3}}{4} f^{\{1}g_{\{ \dot 0} {(D\sigma)^0}_{\dot 0} {(D\sigma)^{0\}}}_{\dot 0} h_{\dot 0 \}} + \frac{\sqrt{3}}{4} f^{\{1}g_{\{ \dot 1} {(D\sigma)^0}_{\dot 1} {(D\sigma)^{0\}}}_{\dot 1} h_{\dot 1 \}} \nonumber\\ & + \frac{\sqrt{3}}{2} f^{\{1}g_{\{ \dot 1} {(D\sigma)^1}_{\dot 1} {(D\sigma)^{1\}}}_{\dot 0} h_{\dot 0 \}}, \nonumber\\
\mathcal{O}_{DD7}^{(8)} =& \frac{3}{2} f^{\{1}g_{\{ \dot 1} {(D\sigma)^0}_{\dot 1} {(D\sigma)^{0\}}}_{\dot 0} h_{\dot 0 \}} - \frac{1}{4} f^{\{1}g_{\{ \dot 0} {(D\sigma)^1}_{\dot 0} {(D\sigma)^{1\}}}_{\dot 0} h_{\dot 0 \}} \nonumber\\ & - \frac{1}{4} f^{\{1}g_{\{ \dot 1} {(D\sigma)^1}_{\dot 1} {(D\sigma)^{1\}}}_{\dot 1} h_{\dot 1 \}}.
\end{align}
\normalsize
Once more, $\mathcal{O}_{DD8}^{(i)}$ ($\mathcal{O}_{DD9}^{(i)}$) are obtained after exchange of the index on the first quark with that on the second (third) quark.

The next operator multiplets belong to the irreducible representation $\tau^{\underline{12}}_1$:
\scriptsize
\begin{align}
\mathcal{O}_{DD10}^{(1)} =& -\frac{3}{2} f^{\{1}g_{\{ \dot 1} {(D\sigma)^0}_{\dot 0} {(D\sigma)^{0\}}}_{\dot 0} h_{\dot 0 \}} - \frac{1}{2} f^{\{1}g_{\{ \dot 1} {(D\sigma)^1}_{\dot 1} {(D\sigma)^{1\}}}_{\dot 1} h_{\dot 0 \}}, \nonumber\\
\mathcal{O}_{DD10}^{(2)} =& \frac{1}{2\sqrt{2}}  \left( f^{\{1}g_{\{ \dot 0} {(D\sigma)^1}_{\dot 0} {(D\sigma)^{1\}}}_{\dot 0} h_{\dot 0 \}} - f^{\{1}g_{\{ \dot 1} {(D\sigma)^1}_{\dot 1} {(D\sigma)^{1\}}}_{\dot 1} h_{\dot 1 \}} \right), \nonumber\\
\mathcal{O}_{DD10}^{(3)} =& \frac{3}{2} f^{\{1}g_{\{ \dot 1} {(D\sigma)^0}_{\dot 1} {(D\sigma)^{0\}}}_{\dot 1} h_{\dot 0 \}} + \frac{1}{2} f^{\{1}g_{\{ \dot 1} {(D\sigma)^1}_{\dot 0} {(D\sigma)^{1\}}}_{\dot 0} h_{\dot 0 \}}, \nonumber\\
\mathcal{O}_{DD10}^{(4)} =& \frac{1}{2} f^{\{0}g_{\{ \dot 1} {(D\sigma)^0}_{\dot 1} {(D\sigma)^{0\}}}_{\dot 1} h_{\dot 0 \}} + \frac{3}{2} f^{\{1}g_{\{ \dot 1} {(D\sigma)^1}_{\dot 0} {(D\sigma)^{0\}}}_{\dot 0} h_{\dot 0 \}}, \nonumber\\
\mathcal{O}_{DD10}^{(5)} =& \frac{1}{2\sqrt{2}} \left( f^{\{0}g_{\{ \dot 1} {(D\sigma)^0}_{\dot 1} {(D\sigma)^{0\}}}_{\dot 1} h_{\dot 1 \}} - f^{\{0}g_{\{ \dot 0} {(D\sigma)^0}_{\dot 0} {(D\sigma)^{0\}}}_{\dot 0} h_{\dot 0 \}} \right), \nonumber\\
\mathcal{O}_{DD10}^{(6)} =& -\frac{1}{2} f^{\{0}g_{\{ \dot 1} {(D\sigma)^0}_{\dot 0} {(D\sigma)^{0\}}}_{\dot 0} h_{\dot 0 \}} - \frac{3}{2} f^{\{1}g_{\{ \dot 1} {(D\sigma)^1}_{\dot 1} {(D\sigma)^{0\}}}_{\dot 1} h_{\dot 0 \}}, \nonumber\\
\mathcal{O}_{DD10}^{(7)} =& -\frac{3}{2} f_{\{\dot 1}g^{\{1} {(D\sigma)^0}_{\dot 0} {(D\sigma)^0}_{\dot 0 \}} h^{0\}} - \frac{1}{2} f_{\{\dot 1}g^{\{ 1} {(D\sigma)^1}_{\dot 1} {(D\sigma)^1}_{\dot 1 \}} h^{0\}}, \nonumber\\
\mathcal{O}_{DD10}^{(8)} =& \frac{1}{2\sqrt{2}} \left( f_{\{\dot 1}g^{\{ 0} {(D\sigma)^0}_{\dot 1} {(D\sigma)^0}_{\dot 1 \}} h^{0\}} - f_{\{\dot 1}g^{\{ 1} {(D\sigma)^1}_{\dot 1} {(D\sigma)^1}_{\dot 1 \}} h^{1\}} \right), \nonumber\\
\mathcal{O}_{DD10}^{(9)} =& \frac{3}{2} f_{\{\dot 1}g^{\{1} {(D\sigma)^1}_{\dot 0} {(D\sigma)^1}_{\dot 0 \}} h^{0\}} + \frac{1}{2} f_{\{\dot 1}g^{\{ 1} {(D\sigma)^0}_{\dot 1} {(D\sigma)^0}_{\dot 1 \}} h^{0\}}, \nonumber\\
\mathcal{O}_{DD10}^{(10)} =& \frac{1}{2} f_{\{\dot 0}g^{\{1} {(D\sigma)^1}_{\dot 0} {(D\sigma)^1}_{\dot 0 \}} h^{0\}} + \frac{3}{2} f_{\{\dot 1}g^{\{1} {(D\sigma)^0}_{\dot 1} {(D\sigma)^0}_{\dot 0 \}} h^{0\}}, \nonumber\\
\mathcal{O}_{DD10}^{(11)} =& \frac{1}{2\sqrt{2}} \left( f_{\{\dot 0}g^{\{1} {(D\sigma)^1}_{\dot 0} {(D\sigma)^1}_{\dot 0 \}} h^{1\}} - f_{\{\dot 0}g^{\{0} {(D\sigma)^0}_{\dot 0} {(D\sigma)^0}_{\dot 0 \}} h^{0\}} \right), \nonumber\\
\mathcal{O}_{DD10}^{(12)} =& -\frac{1}{2} f_{\{\dot 0}g^{\{1} {(D\sigma)^0}_{\dot 0} {(D\sigma)^0}_{\dot 0 \}} h^{0\}} - \frac{3}{2} f_{\{\dot 1}g^{\{1} {(D\sigma)^1}_{\dot 1} {(D\sigma)^1}_{\dot 0 \}} h^{0\}}.
\end{align}
\normalsize
The operators $\mathcal{O}_{DD11}^{(i)}$ ($\mathcal{O}_{DD12}^{(i)}$) result upon interchanging the indices on $f$ and $g$ ($h$).
\scriptsize
\begin{align}
\mathcal{O}_{DD13}^{(1)} =& \frac{\sqrt{5}}{4\sqrt{2}} \left(-2\cdot f^{\{ 1} g^{1} {(D\sigma)^0}_{\{ \dot 0} {(D\sigma)^0}_{\dot 0 \}} h^{0 \}} \right. \nonumber\\ & \left. - f^{\{ 0} g^{0} {(D\sigma)^0}_{\{ \dot 1} {(D\sigma)^0}_{\dot 1 \}} h^{0 \}} -f^{\{ 1} g^{1} {(D\sigma)^1}_{\{ \dot 1} {(D\sigma)^1}_{\dot 1 \}} h^{0 \}} \right), \nonumber \\
\mathcal{O}_{DD13}^{(2)} =& +\sqrt{5} f^{\{ 1} g^{1} {(D\sigma)^0}_{\{ \dot 1} {(D\sigma)^0}_{\dot 0 \}} h^{0 \}}, \nonumber\\
\mathcal{O}_{DD13}^{(3)} =& \frac{\sqrt{5}}{4\sqrt{2}} \left( -f^{\{ 0} g^{0} {(D\sigma)^0}_{\{ \dot 0} {(D\sigma)^0}_{\dot 0 \}} h^{0 \}} \right. \nonumber\\ &\left. - f^{\{ 1} g^{1} {(D\sigma)^1}_{\{ \dot 0} {(D\sigma)^1}_{\dot 0 \}} h^{0 \}} -2 \cdot f^{\{ 1} g^{1} {(D\sigma)^0}_{\{ \dot 1} {(D\sigma)^0}_{\dot 1 \}} h^{0 \}} \right), \nonumber\\
\mathcal{O}_{DD13}^{(4)} =& \frac{\sqrt{5}}{4\sqrt{2}} \left( - 2\cdot f^{\{ 1} g^{1} {(D\sigma)^1}_{\{ \dot 0} {(D\sigma)^0}_{\dot 0 \}} h^{0 \}} \right. \nonumber\\ &  \left. -f^{\{ 1} g^{0} {(D\sigma)^0}_{\{ \dot 1} {(D\sigma)^0}_{\dot 1 \}} h^{0 \}} - f^{\{ 1} g^{1} {(D\sigma)^1}_{\{ \dot 1} {(D\sigma)^1}_{\dot 1 \}} h^{1 \}} \right), \nonumber\\
\mathcal{O}_{DD13}^{(5)} =& \sqrt{5} f^{\{ 1} g^{1} {(D\sigma)^1}_{\{ \dot 1} {(D\sigma)^0}_{\dot 0 \}} h^{0 \}}, \nonumber\\
\mathcal{O}_{DD13}^{(6)} =& \frac{\sqrt{5}}{4\sqrt{2}} \left( - \cdot f^{\{ 1} g^{0} {(D\sigma)^0}_{\{ \dot 0} {(D\sigma)^0}_{\dot 0 \}} h^{0 \}} \right. \nonumber\\ & \left. - f^{\{ 1} g^{1} {(D\sigma)^1}_{\{ \dot 0} {(D\sigma)^1}_{\dot 0 \}} h^{1 \}}  - 2\cdot f^{\{ 1} g^{1} {(D\sigma)^1}_{\{ \dot 1} {(D\sigma)^0}_{\dot 1 \}} h^{0 \}} \right), \nonumber\\
\mathcal{O}_{DD13}^{(7)} =& \frac{\sqrt{5}}{4\sqrt{2}} \left( - 2\cdot f_{\{ \dot 1} g_{\dot 1} {(D\sigma)^{\{0}}_{\dot 0} {(D\sigma)^{0 \}}}_{\dot 0} h_{\dot 0 \}} \right. \nonumber\\ &\left.- f_{\{ \dot 0} g_{\dot 0} {(D\sigma)^{\{1}}_{\dot 0} {(D\sigma)^{1\}}}_{\dot 0} h_{\dot 0 \}} - f_{\{ \dot 1} g_{\dot 1} {(D\sigma)^{\{1}}_{\dot 1} {(D\sigma)^{1\}}}_{\dot 1} h_{\dot 0 \}} \right), \nonumber\\
\mathcal{O}_{DD13}^{(8)} =& \sqrt{5} f_{\{ \dot 1} g_{\dot 1} {(D\sigma)^{\{1}}_{\dot 0} {(D\sigma)^{0 \}}}_{\dot 0} h_{\dot 0 \}}, \nonumber\\
\mathcal{O}_{DD13}^{(9)} =& \frac{\sqrt{5}}{4\sqrt{2}} \left( -f_{\{ \dot 0} g_{\dot 0} {(D\sigma)^{\{0}}_{\dot 0} {(D\sigma)^{0 \}}}_{\dot 0} h_{\dot 0 \}} \right. \nonumber\\ & \left. - f_{\{ \dot 1} g_{\dot 1} {(D\sigma)^{\{0}}_{\dot 1} {(D\sigma)^{0 \}}}_{\dot 1} h_{\dot 0 \}} - 2\cdot f_{\{ \dot 1} g_{\dot 1} {(D\sigma)^{\{1}}_{\dot 0} {(D\sigma)^{1\}}}_{\dot 0} h_{\dot 0 \}} \right), \nonumber\\
\mathcal{O}_{DD13}^{(10)} =& \frac{\sqrt{5}}{4\sqrt{2}} \left( -2\cdot f_{\{ \dot 1} g_{\dot 1} {(D\sigma)^{\{0}}_{\dot 1} {(D\sigma)^{0 \}}}_{\dot 0} h_{\dot 0 \}} \right. \nonumber\\ &\left. - f_{\{ \dot 1} g_{\dot 0} {(D\sigma)^{\{1}}_{\dot 0} {(D\sigma)^{1\}}}_{\dot 0} h_{\dot 0 \}}  - f_{\{ \dot 1} g_{\dot 1} {(D\sigma)^{\{1}}_{\dot1} {(D\sigma)^{1\}}}_{\dot1} h_{\dot1 \}} \right), \nonumber\\
\mathcal{O}_{DD13}^{(11)} =& +\sqrt{5} f_{\{ \dot 1} g_{\dot 1} {(D\sigma)^{\{1}}_{\dot 1} {(D\sigma)^{0 \}}}_{\dot 0} h_{\dot 0 \}}, \nonumber\\
\mathcal{O}_{DD13}^{(12)} =& \frac{\sqrt{5}}{4\sqrt{2}} \left( - f_{\{ \dot 1} g_{\dot 0} {(D\sigma)^{\{0}}_{\dot 0} {(D\sigma)^{0 \}}}_{\dot 0} h_{\dot 0 \}} \right. \nonumber\\ & \left.- f_{\{ \dot 1} g_{\dot 1} {(D\sigma)^{\{0}}_{\dot 1} {(D\sigma)^{0 \}}}_{\dot 1} h_{\dot 1 \}} - 2\cdot f_{\{ \dot 1} g_{\dot 1} {(D\sigma)^{\{1}}_{\dot 1} {(D\sigma)^{1\}}}_{\dot 0} h_{\dot 0 \}} \right).
\end{align}
\normalsize
Furthermore there are the five $\tau^{\underline{12}}_2$ multiplets. We start with the multiplet $DD14$:
\scriptsize
\begin{align}
\mathcal{O}_{DD14}^{(1)} =& \frac{\sqrt{3}}{2} \left( f^{\{1}g_{\{ \dot 1} {(D\sigma)^1}_{\dot 1} {(D\sigma)^{0\}}}_{\dot 1} h_{\dot 0 \}} - f^{\{0}g_{\{ \dot 1} {(D\sigma)^0}_{\dot 0} {(D\sigma)^{0\}}}_{\dot 0} h_{\dot 0 \}} \right), \nonumber\\
\mathcal{O}_{DD14}^{(2)} =& \frac{\sqrt{3}}{2} \left( f^{\{1}g_{\{ \dot 1} {(D\sigma)^0}_{\dot 0} {(D\sigma)^{0\}}}_{\dot 0} h_{\dot 0 \}} - f^{\{1}g_{\{ \dot 1} {(D\sigma)^1}_{\dot 1} {(D\sigma)^{1\}}}_{\dot 1} h_{\dot 0 \}} \right), \nonumber\\
\mathcal{O}_{DD14}^{(3)} =& \frac{\sqrt{3}}{2} \left( f^{\{1}g_{\{ \dot 1} {(D\sigma)^1}_{\dot 0} {(D\sigma)^{0\}}}_{\dot 0} h_{\dot 0 \}} - f^{\{0}g_{\{ \dot 1} {(D\sigma)^0}_{\dot 1} {(D\sigma)^{0\}}}_{\dot 1} h_{\dot 0 \}} \right), \nonumber\\
\mathcal{O}_{DD14}^{(4)} =& \frac{\sqrt{3}}{2} \left( f^{\{1}g_{\{ \dot 1} {(D\sigma)^0}_{\dot 1} {(D\sigma)^{0\}}}_{\dot 1} h_{\dot 0 \}} - f^{\{1}g_{\{ \dot 1} {(D\sigma)^1}_{\dot 0} {(D\sigma)^{1\}}}_{\dot 0} h_{\dot 0 \}} \right), \nonumber\\
\mathcal{O}_{DD14}^{(5)} =& \frac{\sqrt{3}}{2 \sqrt{2}} \left( f^{\{1}g_{\{ \dot 0} {(D\sigma)^1}_{\dot 0} {(D\sigma)^{0\}}}_{\dot 0} h_{\dot 0 \}} - f^{\{1}g_{\{ \dot 1} {(D\sigma)^1}_{\dot 1} {(D\sigma)^{0\}}}_{\dot 1} h_{\dot 1 \}} \right), \nonumber\\
\mathcal{O}_{DD14}^{(6)} =& \frac{\sqrt{3}}{2 \sqrt{2}} \left( f^{\{1}g_{\{ \dot 0} {(D\sigma)^0}_{\dot 0} {(D\sigma)^{0\}}}_{\dot 0} h_{\dot 0 \}} - f^{\{1}g_{\{ \dot 1} {(D\sigma)^0}_{\dot 1} {(D\sigma)^{0\}}}_{\dot 1} h_{\dot 1 \}} \right), \nonumber\\
\mathcal{O}_{DD14}^{(7)} =& \frac{\sqrt{3}}{2} \left( f_{\{\dot 0}g^{\{1} {(D\sigma)^0}_{\dot 0} {(D\sigma)^0}_{\dot 0 \}} h^{0\}} - f_{\{\dot 1}g^{\{1} {(D\sigma)^1}_{\dot 1} {(D\sigma)^1}_{\dot 0 \}} h^{0\}} \right), \nonumber\\
\mathcal{O}_{DD14}^{(8)} =& \frac{\sqrt{3}}{2} \left( f_{\{\dot 1}g^{\{ 1} {(D\sigma)^1}_{\dot 1} {(D\sigma)^1}_{\dot 1 \}} h^{0\}} - f_{\{\dot 1}g^{\{1} {(D\sigma)^0}_{\dot 0} {(D\sigma)^0}_{\dot 0 \}} h^{0\}} \right), \nonumber\\
\mathcal{O}_{DD14}^{(9)} =& \frac{\sqrt{3}}{2} \left( f_{\{\dot 0}g^{\{1} {(D\sigma)^1}_{\dot 0} {(D\sigma)^1}_{\dot 0 \}} h^{0\}} - f_{\{\dot 1}g^{\{1} {(D\sigma)^0}_{\dot 1} {(D\sigma)^0}_{\dot 0 \}} h^{0\}} \right), \nonumber\\
\mathcal{O}_{DD14}^{(10)} =& \frac{\sqrt{3}}{2} \left( f_{\{\dot 1}g^{\{ 1} {(D\sigma)^0}_{\dot 1} {(D\sigma)^0}_{\dot 1 \}} h^{0\}} - f_{\{\dot 1}g^{\{1} {(D\sigma)^1}_{\dot 0} {(D\sigma)^1}_{\dot 0 \}} h^{0\}} \right), \nonumber\\
\mathcal{O}_{DD14}^{(11)} =& \frac{\sqrt{3}}{2 \sqrt{2}} \left( f_{\{\dot 1}g^{\{1} {(D\sigma)^1}_{\dot 1} {(D\sigma)^1}_{\dot 0 \}} h^{1\}} - f_{\{\dot 1}g^{\{0} {(D\sigma)^0}_{\dot 1} {(D\sigma)^0}_{\dot 0 \}} h^{0\}} \right), \nonumber\\
\mathcal{O}_{DD14}^{(12)} =& \frac{\sqrt{3}}{2 \sqrt{2}} \left( f_{\{\dot 1}g^{\{1} {(D\sigma)^1}_{\dot 0} {(D\sigma)^1}_{\dot 0 \}} h^{1\}} - f_{\{\dot 1}g^{\{0} {(D\sigma)^0}_{\dot 0} {(D\sigma)^0}_{\dot 0 \}} h^{0\}} \right).
\end{align}
\normalsize
Again, the chirality partners $\mathcal{O}_{DD15}^{(i)}$ ($\mathcal{O}_{DD16}^{(i)}$) are generated by exchange of the index on the first with that on the second (third) quark field.

Finally, two equivalent multiplets exist that originate from one $\overline{O_4}$ irreducible multiplet. They were separated using the projectors $\tilde P^\alpha_{lk}$ introduced in eq. (\ref{equiv_projector}):
\scriptsize
\begin{align}
\mathcal{O}_{DD17}^{(1)} =& \frac{5i}{4\sqrt{6}} \left( 2 \cdot f^{\{ 1} g^{1} {(D\sigma)^1}_{\{ \dot 1} {(D\sigma)^0}_{\dot 1 \}} h^{0 \}} \right. \nonumber\\ 
                          & \left. - \frac{3}{5} f^{\{ 1} g^{1} {(D\sigma)^1}_{\{ \dot 0} {(D\sigma)^1}_{\dot 0 \}} h^{1 \}} + f^{\{ 1} g^{0} {(D\sigma)^0}_{\{ \dot 0} {(D\sigma)^0}_{\dot 0 \}} h^{0 \}} \right), \nonumber\\
\mathcal{O}_{DD17}^{(2)} =& \frac{5i}{4\sqrt{6}} \left( 2 \cdot f^{\{ 1} g^{1} {(D\sigma)^0}_{\{ \dot 0} {(D\sigma)^0}_{\dot 0 \}} h^{0 \}} \right. \nonumber\\ 
                          & \left. - \frac{3}{5} f^{\{ 0} g^{0} {(D\sigma)^0}_{\{ \dot 1} {(D\sigma)^0}_{\dot 1 \}} h^{0 \}} + f^{\{ 1} g^{1} {(D\sigma)^1}_{\{ \dot 1} {(D\sigma)^1}_{\dot 1 \}} h^{0 \}} \right), \nonumber\\
\mathcal{O}_{DD17}^{(3)} =& \frac{5i}{4\sqrt{6}} \left( - 2\cdot f^{\{ 1} g^{1} {(D\sigma)^1}_{\{ \dot 0} {(D\sigma)^0}_{\dot 0 \}} h^{0 \}} \right. \nonumber\\ 
                          & \left. - f^{\{ 1} g^{0} {(D\sigma)^0}_{\{ \dot 1} {(D\sigma)^0}_{\dot 1 \}} h^{0 \}} + \frac{3}{5} f^{\{ 1} g^{1} {(D\sigma)^1}_{\{ \dot 1} {(D\sigma)^1}_{\dot 1 \}} h^{1 \}} \right), \nonumber\\
\mathcal{O}_{DD17}^{(4)} =& \frac{5i}{4\sqrt{6}} \left( \frac{3}{5} f^{\{ 0} g^{0} {(D\sigma)^0}_{\{ \dot 0} {(D\sigma)^0}_{\dot 0 \}} h^{0 \}} \right. \nonumber\\ 
                          & \left. - f^{\{ 1} g^{1} {(D\sigma)^1}_{\{ \dot 0} {(D\sigma)^1}_{\dot 0 \}} h^{0 \}} - 2 \cdot f^{\{ 1} g^{1} {(D\sigma)^0}_{\{ \dot 1} {(D\sigma)^0}_{\dot 1 \}} h^{0 \}} \right), \nonumber\\
\mathcal{O}_{DD17}^{(5)} =& \frac{-i}{2\sqrt{3}} \left( 5\cdot f^{\{ 1} g^{0} {(D\sigma)^0}_{\{ \dot 1} {(D\sigma)^0}_{\dot 0 \}} h^{0 \}} \right. \nonumber\\ 
                          & \left. + f^{\{ 1} g^{1} {(D\sigma)^1}_{\{ \dot 1} {(D\sigma)^1}_{\dot 0 \}} h^{1 \}} \right), \nonumber\\
\mathcal{O}_{DD17}^{(6)} =& \frac{i}{2\sqrt{3}} \left( f^{\{ 0} g^{0} {(D\sigma)^0}_{\{ \dot 1} {(D\sigma)^0}_{\dot 0 \}} h^{0 \}} \right. \nonumber\\
                          & \left. + 5 \cdot f^{\{ 1} g^{1} {(D\sigma)^1}_{\{ \dot 1} {(D\sigma)^1}_{\dot 0 \}} h^{0 \}} \right), \nonumber\\
\mathcal{O}_{DD17}^{(7)} =& \frac{5i}{4\sqrt{6}} \left( - 2 \cdot f_{\{ \dot 1} g_{\dot 1} {(D\sigma)^{\{1}}_{\dot 1} {(D\sigma)^{1\}}}_{\dot 0} h_{\dot 0 \}} \right. \nonumber\\
                          &\left. + \frac{3}{5} f_{\{ \dot 1} g_{\dot 1} {(D\sigma)^{\{0}}_{\dot 1} {(D\sigma)^{0 \}}}_{\dot 1} h_{\dot 1 \}} -f_{\{ \dot 1} g_{\dot 0} {(D\sigma)^{\{0}}_{\dot 0} {(D\sigma)^{0 \}}}_{\dot 0} h_{\dot 0 \}} \right), \nonumber\\
\mathcal{O}_{DD17}^{(8)} =& \frac{5i}{4\sqrt{6}} \left( -2\cdot f_{\{ \dot 1} g_{\dot 1} {(D\sigma)^{\{0}}_{\dot 0} {(D\sigma)^{0 \}}}_{\dot 0} h_{\dot 0 \}} \right. \nonumber\\
                          & \left. + \frac{3}{5} f_{\{ \dot 0} g_{\dot 0} {(D\sigma)^{\{1}}_{\dot 0} {(D\sigma)^{1\}}}_{\dot 0} h_{\dot 0 \}} - f_{\{ \dot 1} g_{\dot 1} {(D\sigma)^{\{1}}_{\dot 1} {(D\sigma)^{1\}}}_{\dot 1} h_{\dot 0 \}} \right), \nonumber\\
\mathcal{O}_{DD17}^{(9)} =& \frac{5i}{4\sqrt{6}} \left( 2\cdot f_{\{ \dot 1} g_{\dot 1} {(D\sigma)^{\{0}}_{\dot 1} {(D\sigma)^{0 \}}}_{\dot 0} h_{\dot 0 \}} \right. \nonumber\\
                          & \left. + f_{\{ \dot 1} g_{\dot 0} {(D\sigma)^{\{1}}_{\dot 0} {(D\sigma)^{1\}}}_{\dot 0} h_{\dot 0 \}} - \frac{3}{5} f_{\{ \dot 1} g_{\dot 1} {(D\sigma)^{\{1}}_{\dot1} {(D\sigma)^{1\}}}_{\dot1} h_{\dot1 \}} \right), \nonumber\\
\mathcal{O}_{DD17}^{(10)} =& \frac{5i}{4\sqrt{6}} \left( -\frac{3}{5} f_{\{ \dot 0} g_{\dot 0} {(D\sigma)^{\{0}}_{\dot 0} {(D\sigma)^{0 \}}}_{\dot 0} h_{\dot 0 \}} \right. \nonumber\\ 
                           & \left. + f_{\{ \dot 1} g_{\dot 1} {(D\sigma)^{\{0}}_{\dot 1} {(D\sigma)^{0 \}}}_{\dot 1} h_{\dot 0 \}} + 2\cdot f_{\{ \dot 1} g_{\dot 1} {(D\sigma)^{\{1}}_{\dot 0} {(D\sigma)^{1\}}}_{\dot 0} h_{\dot 0 \}} \right), \nonumber\\
\mathcal{O}_{DD17}^{(11)} =& \frac{i}{2\sqrt{3}} \left( 5\cdot f_{\{ \dot 1} g_{\dot 0} {(D\sigma)^{\{1}}_{\dot 0} {(D\sigma)^{0 \}}}_{\dot 0} h_{\dot 0 \}} \right. \nonumber\\
                           & \left. + f_{\{ \dot 1} g_{\dot 1} {(D\sigma)^{\{1}}_{\dot 1} {(D\sigma)^{0 \}}}_{\dot 1} h_{\dot 1 \}} \right), \nonumber\\
\mathcal{O}_{DD17}^{(12)} =& \frac{-i}{2\sqrt{3}} \left( f_{\{ \dot 0} g_{\dot 0} {(D\sigma)^{\{1}}_{\dot 0} {(D\sigma)^{0 \}}}_{\dot 0} h_{\dot 0 \}} \right. \nonumber\\
                           & \left. + 5\cdot f_{\{ \dot 1} g_{\dot 1} {(D\sigma)^{\{1}}_{\dot 1} {(D\sigma)^{0 \}}}_{\dot 1} h_{\dot 0 \}} \right)
\end{align}
\normalsize
and
\scriptsize
\begin{align}
\mathcal{O}_{DD18}^{(1)} =& \sqrt{\frac{5}{6}} \left( f^{\{ 1} g^{0} {(D\sigma)^0}_{\{ \dot 0} {(D\sigma)^0}_{\dot 0 \}} h^{0 \}} - f^{\{ 1} g^{1} {(D\sigma)^1}_{\{ \dot 1} {(D\sigma)^0}_{\dot 1 \}} h^{0 \}} \right), \nonumber\\
\mathcal{O}_{DD18}^{(2)} =& \sqrt{\frac{5}{6}} \left( f^{\{ 1} g^{1} {(D\sigma)^1}_{\{ \dot 1} {(D\sigma)^1}_{\dot 1 \}} h^{0 \}} - f^{\{ 1} g^{1} {(D\sigma)^0}_{\{ \dot 0} {(D\sigma)^0}_{\dot 0 \}} h^{0 \}} \right), \nonumber\\
\mathcal{O}_{DD18}^{(3)} =& \sqrt{\frac{5}{6}} \left( f^{\{ 1} g^{1} {(D\sigma)^1}_{\{ \dot 0} {(D\sigma)^0}_{\dot 0 \}} h^{0 \}} - f^{\{ 1} g^{0} {(D\sigma)^0}_{\{ \dot 1} {(D\sigma)^0}_{\dot 1 \}} h^{0 \}} \right), \nonumber\\
\mathcal{O}_{DD18}^{(4)} =& \sqrt{\frac{5}{6}} \left( f^{\{ 1} g^{1} {(D\sigma)^0}_{\{ \dot 1} {(D\sigma)^0}_{\dot 1 \}} h^{0 \}} - f^{\{ 1} g^{1} {(D\sigma)^1}_{\{ \dot 0} {(D\sigma)^1}_{\dot 0 \}} h^{0 \}} \right), \nonumber\\
\mathcal{O}_{DD18}^{(5)} =& \frac{\sqrt{5}}{2\sqrt{3}} \left( f^{\{ 1} g^{1} {(D\sigma)^1}_{\{ \dot 1} {(D\sigma)^1}_{\dot 0 \}} h^{1 \}} - f^{\{ 1} g^{0} {(D\sigma)^0}_{\{ \dot 1} {(D\sigma)^0}_{\dot 0 \}} h^{0 \}} \right), \nonumber\\
\mathcal{O}_{DD18}^{(6)} =& \frac{\sqrt{5}}{2\sqrt{3}} \left( f^{\{ 1} g^{1} {(D\sigma)^1}_{\{ \dot 1} {(D\sigma)^1}_{\dot 0 \}} h^{0 \}} - f^{\{ 0} g^{0} {(D\sigma)^0}_{\{ \dot 1} {(D\sigma)^0}_{\dot 0 \}} h^{0 \}} \right), \nonumber\\
\mathcal{O}_{DD18}^{(7)} =& \sqrt{\frac{5}{6}} \left( f_{\{ \dot 1} g_{\dot 1} {(D\sigma)^{\{1}}_{\dot 1} {(D\sigma)^{1\}}}_{\dot 0} h_{\dot 0 \}} - f_{\{ \dot 1} g_{\dot 0} {(D\sigma)^{\{0}}_{\dot 0} {(D\sigma)^{0 \}}}_{\dot 0} h_{\dot 0 \}} \right), \nonumber\\
\mathcal{O}_{DD18}^{(8)} =& \sqrt{\frac{5}{6}} \left( f_{\{ \dot 1} g_{\dot 1} {(D\sigma)^{\{0}}_{\dot 0} {(D\sigma)^{0 \}}}_{\dot 0} h_{\dot 0 \}} - f_{\{ \dot 1} g_{\dot 1} {(D\sigma)^{\{1}}_{\dot 1} {(D\sigma)^{1\}}}_{\dot 1} h_{\dot 0 \}} \right), \nonumber\\
\mathcal{O}_{DD18}^{(9)} =& \sqrt{\frac{5}{6}} \left( f_{\{ \dot 1} g_{\dot 0} {(D\sigma)^{\{1}}_{\dot 0} {(D\sigma)^{1\}}}_{\dot 0} h_{\dot 0 \}} - f_{\{ \dot 1} g_{\dot 1} {(D\sigma)^{\{0}}_{\dot 1} {(D\sigma)^{0 \}}}_{\dot 0} h_{\dot 0 \}} \right), \nonumber\\
\mathcal{O}_{DD18}^{(10)} =& \sqrt{\frac{5}{6}} \left( f_{\{ \dot 1} g_{\dot 1} {(D\sigma)^{\{0}}_{\dot 1} {(D\sigma)^{0 \}}}_{\dot 1} h_{\dot 0 \}} - f_{\{ \dot 1} g_{\dot 1} {(D\sigma)^{\{1}}_{\dot 0} {(D\sigma)^{1\}}}_{\dot 0} h_{\dot 0 \}} \right), \nonumber\\
\mathcal{O}_{DD18}^{(11)} =& \frac{\sqrt{5}}{2\sqrt{3}} \left( f_{\{ \dot 1} g_{\dot 0} {(D\sigma)^{\{1}}_{\dot 0} {(D\sigma)^{0 \}}}_{\dot 0} h_{\dot 0 \}} - f_{\{ \dot 1} g_{\dot 1} {(D\sigma)^{\{1}}_{\dot 1} {(D\sigma)^{0 \}}}_{\dot 1} h_{\dot 1 \}} \right), \nonumber\\
\mathcal{O}_{DD18}^{(12)} =& \frac{\sqrt{5}}{2\sqrt{3}} \left( f_{\{ \dot 0} g_{\dot 0} {(D\sigma)^{\{1}}_{\dot 0} {(D\sigma)^{0 \}}}_{\dot 0} h_{\dot 0 \}} - f_{\{ \dot 1} g_{\dot 1} {(D\sigma)^{\{1}}_{\dot 1} {(D\sigma)^{0 \}}}_{\dot 1} h_{\dot 0 \}} \right).
\end{align}
\normalsize

\section{Isospin induced Identities}
\label{App:5}
In this Appendix we summarize our results for isospin $1/2$ symmetrized three-quark operators. Exploiting identities between them we arrive at a minimal independent set of multiplets of three-quark operators with leading twist, up to two derivatives and isospin $1/2$.

\subsection{Operators without Derivatives}
\label{App:6}
In a first step all mixed symmetric isospin operators can be reexpressed in terms of mixed antisymmetric isospin operators:
\begin{align}
\mathcal{O}^{(i),MS}_{1} &= +\frac{1}{\sqrt{3}} \cdot \mathcal{O}^{(i),MA}_{1}, \nonumber\\
\mathcal{O}^{(i),MS}_{2} &= -\frac{2}{\sqrt{3}} \cdot \mathcal{O}^{(i),MA}_{1}, \nonumber\\
\mathcal{O}^{(i),MS}_{3} &= +\sqrt{3} \cdot \mathcal{O}^{(i),MA}_{3}, \nonumber\\
\mathcal{O}^{(i),MS}_{4} &= +\sqrt{3} \cdot \mathcal{O}^{(i),MA}_{3}, \nonumber\\
\mathcal{O}^{(i),MS}_{5} &= 0, \nonumber\\
\mathcal{O}^{(i),MS}_{6} &= 0, \nonumber\\
\mathcal{O}^{(i),MS}_{7} &= -\frac{1}{\sqrt{3}} \cdot \mathcal{O}^{(i),MA}_{7}, \nonumber\\
\mathcal{O}^{(i),MS}_{8} &= -\frac{1}{\sqrt{3}} \cdot \mathcal{O}^{(i),MA}_{7}, \nonumber\\
\mathcal{O}^{(i),MS}_{9} &= +\frac{2}{\sqrt{3}} \cdot \mathcal{O}^{(i),MA}_{7}.
\end{align}
Then one can derive a set of identities among the mixed antisymmetric operators uncovering further dependencies:
\begin{align}
\mathcal{O}^{(i),MA}_{2} &= 0, \nonumber\\
\mathcal{O}^{(i),MA}_{4} &= -1 \cdot \mathcal{O}^{(i),MA}_{3}, \nonumber\\
\mathcal{O}^{(i),MA}_{5} &= -2 \cdot \mathcal{O}^{(i),MA}_{3}, \nonumber\\
\mathcal{O}^{(i),MA}_{6} &= 0, \nonumber\\
\mathcal{O}^{(i),MA}_{8} &= -1 \cdot \mathcal{O}^{(i),MA}_{7}, \nonumber\\
\mathcal{O}^{(i),MA}_{9} &= 0.
\end{align}
So we can take as a minimal set of linearly independent operators the multiplets $\mathcal{O}^{(i),MA}_{1}$, $\mathcal{O}^{(i),MA}_{3}$ and $\mathcal{O}^{(i),MA}_{7}$. 

\subsection{Operators with one Derivative}
\label{App:7}
We proceed with operators containing one derivative. Also in this case we eliminate the $MS$ in favor of $MA$ operators. One can even derive identities expressing all operators containing one covariant derivative in terms of those operators which carry the derivative on the first quark field:
\begin{align}
\mathcal{O}^{(i),MS}_{f1} &= -\frac{1}{\sqrt{3}} \cdot \mathcal{O}^{(i),MA}_{f1}, \nonumber\\
\mathcal{O}^{(i),MS}_{f2} &= -\frac{1}{\sqrt{3}} \cdot \mathcal{O}^{(i),MA}_{f2}, \nonumber\\
\mathcal{O}^{(i),MS}_{f3} &= +\frac{1}{\sqrt{3}} \cdot \mathcal{O}^{(i),MA}_{f3} - \frac{2}{\sqrt{3}} \cdot \mathcal{O}^{(i),MA}_{f4}, \nonumber\\
\mathcal{O}^{(i),MS}_{f4} &= -\frac{2}{\sqrt{3}} \cdot \mathcal{O}^{(i),MA}_{f3} + \frac{1}{\sqrt{3}} \cdot \mathcal{O}^{(i),MA}_{f4}, \nonumber\\
\mathcal{O}^{(i),MS}_{f5} &= -\frac{1}{\sqrt{3}} \cdot \mathcal{O}^{(i),MA}_{f5}, \nonumber\\
\mathcal{O}^{(i),MS}_{f6} &= +\frac{1}{\sqrt{3}} \cdot \mathcal{O}^{(i),MA}_{f6} - \frac{2}{\sqrt{3}} \cdot \mathcal{O}^{(i),MA}_{f7}, \nonumber\\
\mathcal{O}^{(i),MS}_{f7} &= -\frac{2}{\sqrt{3}} \cdot \mathcal{O}^{(i),MA}_{f6} + \frac{1}{\sqrt{3}} \cdot \mathcal{O}^{(i),MA}_{f7}, \nonumber\\
\mathcal{O}^{(i),MS}_{f8} &= -\frac{1}{\sqrt{3}} \cdot \mathcal{O}^{(i),MA}_{f8},
\end{align}
\begin{align}
\mathcal{O}^{(i),MS}_{g1} &= -\frac{1}{\sqrt{3}} \cdot \mathcal{O}^{(i),MA}_{f1}, \nonumber\\
\mathcal{O}^{(i),MS}_{g2} &= +\frac{1}{\sqrt{3}} \cdot \mathcal{O}^{(i),MA}_{f3} -\frac{2}{\sqrt{3}} \cdot \mathcal{O}^{(i),MA}_{f4}, \nonumber\\
\mathcal{O}^{(i),MS}_{g3} &= -\frac{1}{\sqrt{3}} \cdot \mathcal{O}^{(i),MA}_{f2}, \nonumber\\
\mathcal{O}^{(i),MS}_{g4} &= -\frac{2}{\sqrt{3}} \cdot \mathcal{O}^{(i),MA}_{f3} +\frac{1}{\sqrt{3}} \cdot \mathcal{O}^{(i),MA}_{f4}, \nonumber\\
\mathcal{O}^{(i),MS}_{g5} &= +\frac{1}{\sqrt{3}} \cdot \mathcal{O}^{(i),MA}_{f6} -\frac{2}{\sqrt{3}} \cdot \mathcal{O}^{(i),MA}_{f7}, \nonumber\\
\mathcal{O}^{(i),MS}_{g6} &= -\frac{1}{\sqrt{3}} \cdot \mathcal{O}^{(i),MA}_{f5}, \nonumber\\
\mathcal{O}^{(i),MS}_{g7} &= -\frac{2}{\sqrt{3}} \cdot \mathcal{O}^{(i),MA}_{f6} +\frac{1}{\sqrt{3}} \cdot \mathcal{O}^{(i),MA}_{f7}, \nonumber\\
\mathcal{O}^{(i),MS}_{g8} &= -\frac{1}{\sqrt{3}} \cdot \mathcal{O}^{(i),MA}_{f8},
\end{align}
\begin{align}
\mathcal{O}^{(i),MS}_{h1} &= +\frac{2}{\sqrt{3}} \cdot \mathcal{O}^{(i),MA}_{f1}, \nonumber\\
\mathcal{O}^{(i),MS}_{h2} &= +\frac{1}{\sqrt{3}} \cdot \mathcal{O}^{(i),MA}_{f3} +\frac{1}{\sqrt{3}} \cdot 
\mathcal{O}^{(i),MA}_{f4}, \nonumber\\
\mathcal{O}^{(i),MS}_{h3} &= +\frac{1}{\sqrt{3}} \cdot \mathcal{O}^{(i),MA}_{f3} +\frac{1}{\sqrt{3}} \cdot \mathcal{O}^{(i),MA}_{f4}, \nonumber\\
\mathcal{O}^{(i),MS}_{h4} &= +\frac{2}{\sqrt{3}} \cdot \mathcal{O}^{(i),MA}_{f2}, \nonumber\\
\mathcal{O}^{(i),MS}_{h5} &= +\frac{1}{\sqrt{3}} \cdot \mathcal{O}^{(i),MA}_{f6} +\frac{1}{\sqrt{3}}, \cdot \mathcal{O}^{(i),MA}_{f7}, \nonumber\\
\mathcal{O}^{(i),MS}_{h6} &= +\frac{1}{\sqrt{3}} \cdot \mathcal{O}^{(i),MA}_{f6} +\frac{1}{\sqrt{3}} \cdot \mathcal{O}^{(i),MA}_{f7}, \nonumber\\
\mathcal{O}^{(i),MS}_{h7} &= +\frac{2}{\sqrt{3}} \cdot \mathcal{O}^{(i),MA}_{f5}, \nonumber\\
\mathcal{O}^{(i),MS}_{h8} &= +\frac{2}{\sqrt{3}} \cdot \mathcal{O}^{(i),MA}_{f8},
\end{align}
\begin{align}
\mathcal{O}^{(i),MA}_{g1} &= - \mathcal{O}^{(i),MA}_{f1}, \nonumber\\
\mathcal{O}^{(i),MA}_{g2} &= - \mathcal{O}^{(i),MA}_{f3}, \nonumber\\
\mathcal{O}^{(i),MA}_{g3} &= - \mathcal{O}^{(i),MA}_{f2}, \nonumber\\
\mathcal{O}^{(i),MA}_{g4} &= - \mathcal{O}^{(i),MA}_{f4}, \nonumber\\
\mathcal{O}^{(i),MA}_{g5} &= - \mathcal{O}^{(i),MA}_{f6}, \nonumber\\
\mathcal{O}^{(i),MA}_{g6} &= - \mathcal{O}^{(i),MA}_{f5}, \nonumber\\
\mathcal{O}^{(i),MA}_{g7} &= - \mathcal{O}^{(i),MA}_{f7}, \nonumber\\
\mathcal{O}^{(i),MA}_{g8} &= - \mathcal{O}^{(i),MA}_{f8},
\end{align}
\begin{align}
\mathcal{O}^{(i),MA}_{h1} &= 0, \nonumber\\
\mathcal{O}^{(i),MA}_{h2} &= - \mathcal{O}^{(i),MA}_{f3} + \mathcal{O}^{(i),MA}_{f4}, \nonumber\\
\mathcal{O}^{(i),MA}_{h3} &= + \mathcal{O}^{(i),MA}_{f3} - \mathcal{O}^{(i),MA}_{f4}, \nonumber\\
\mathcal{O}^{(i),MA}_{h4} &= 0, \nonumber\\
\mathcal{O}^{(i),MA}_{h5} &= - \mathcal{O}^{(i),MA}_{f6} + \mathcal{O}^{(i),MA}_{f7}, \nonumber\\
\mathcal{O}^{(i),MA}_{h6} &= + \mathcal{O}^{(i),MA}_{f6} - \mathcal{O}^{(i),MA}_{f7}, \nonumber\\
\mathcal{O}^{(i),MA}_{h7} &= 0, \nonumber\\
\mathcal{O}^{(i),MA}_{h8} &= 0.
\end{align}

This means that a full set of isospin symmetrized operators with one derivative is given by the irreducible multiplets
\begin{align}
&\mathcal{O}^{(i),MA}_{f1}, &\mathcal{O}^{(i),MA}_{f2},& &\mathcal{O}^{(i),MA}_{f3},& &\mathcal{O}^{(i),MA}_{f4},& \nonumber\\
&\mathcal{O}^{(i),MA}_{f5}, &\mathcal{O}^{(i),MA}_{f6},& &\mathcal{O}^{(i),MA}_{f7},& &\mathcal{O}^{(i),MA}_{f8}.&
\end{align}

\subsection{Operators with two Derivatives}
\label{App:8}
We begin with the $MS$ operators and both derivatives acting on the same quark field. It can be shown that they reduce to those operators which are mixed-antisymmetric in isospin with the derivatives acting on the first quark:
\begin{align}
\mathcal{O}^{(i),MS}_{ff1,4,7,10,14} &= -\frac{1}{\sqrt{3}} \cdot \mathcal{O}^{(i),MA}_{ff1,4,7,10,14}, \nonumber\\
\mathcal{O}^{(i),MS}_{ff2,5,8,11,15} &= +\frac{1}{\sqrt{3}} \cdot \mathcal{O}^{(i),MA}_{ff2,5,8,11,15} - \frac{2}{\sqrt{3}} \cdot \mathcal{O}^{(i),MA}_{ff3,6,9,12,16}, \nonumber\\ 
\mathcal{O}^{(i),MS}_{ff3,6,9,12,16} &= -\frac{2}{\sqrt{3}} \cdot \mathcal{O}^{(i),MA}_{ff2,5,8,11,15} + \frac{1}{\sqrt{3}} \cdot \mathcal{O}^{(i),MA}_{ff3,6,9,12,16}, \nonumber\\ 
\mathcal{O}^{(i),MS}_{ff13,17,18} &= -\frac{1}{\sqrt{3}} \cdot \mathcal{O}^{(i),MA}_{ff13,17,18}.
\end{align}
Let us give a short explanation, how this notation is to be understood. For example,
\begin{align}
\mathcal{O}^{(i),MS}_{ff2,5,8,11,15} &= +\frac{1}{\sqrt{3}} \cdot \mathcal{O}^{(i),MA}_{ff2,5,8,11,15} - \frac{2}{\sqrt{3}} \cdot \mathcal{O}^{(i),MA}_{ff3,6,9,12,16} \nonumber
\end{align}
means:
\begin{align}
\mathcal{O}^{(i),MS}_{ff2} &= +\frac{1}{\sqrt{3}} \cdot \mathcal{O}^{(i),MA}_{ff2} - \frac{2}{\sqrt{3}} \cdot \mathcal{O}^{(i),MA}_{ff3}, \nonumber\\ 
\mathcal{O}^{(i),MS}_{ff5} &= +\frac{1}{\sqrt{3}} \cdot \mathcal{O}^{(i),MA}_{ff5} - \frac{2}{\sqrt{3}} \cdot \mathcal{O}^{(i),MA}_{ff6}, \nonumber\\ 
\vdots \nonumber\\
\mathcal{O}^{(i),MS}_{ff15} &= +\frac{1}{\sqrt{3}} \cdot \mathcal{O}^{(i),MA}_{ff15} - \frac{2}{\sqrt{3}} \cdot \mathcal{O}^{(i),MA}_{ff16}. \nonumber
\end{align}
For mixed symmetric operators with both derivatives acting on the second or third quark we have:
\begin{align}
\mathcal{O}^{(i),MS}_{gg1,4,7,10,14} &= -\frac{1}{\sqrt{3}} \cdot \mathcal{O}^{(i),MA}_{ff2,5,8,11,15} + \frac{2}{\sqrt{3}} \cdot \mathcal{O}^{(i),MA}_{ff3,6,9,12,16}, \nonumber\\ 
\mathcal{O}^{(i),MS}_{gg2,5,8,11,15} &= +\frac{1}{\sqrt{3}} \cdot \mathcal{O}^{(i),MA}_{ff1,4,7,10,14}, \nonumber\\ 
\mathcal{O}^{(i),MS}_{gg3,6,9,12,16} &= -\frac{2}{\sqrt{3}} \cdot \mathcal{O}^{(i),MA}_{ff2,5,8,11,15} + \frac{1}{\sqrt{3}} \cdot \mathcal{O}^{(i),MA}_{ff3,6,9,12,16}, \nonumber\\ 
\mathcal{O}^{(i),MS}_{gg13,17,18} &= -\frac{1}{\sqrt{3}} \cdot \mathcal{O}^{(i),MA}_{ff13,17,18},
\end{align}
\begin{align}
\mathcal{O}^{(i),MS}_{hh1,4,7,10,14} &= -\frac{1}{\sqrt{3}} \cdot \mathcal{O}^{(i),MA}_{ff2,5,8,11,15} - \frac{1}{\sqrt{3}} \cdot \mathcal{O}^{(i),MA}_{ff3,6,9,12,16}, \nonumber\\ 
\mathcal{O}^{(i),MS}_{hh2,5,8,11,15} &= +\frac{1}{\sqrt{3}} \cdot \mathcal{O}^{(i),MA}_{ff2,5,8,11,15} + \frac{1}{\sqrt{3}} \cdot \mathcal{O}^{(i),MA}_{ff3,6,9,12,16}, \nonumber\\ 
\mathcal{O}^{(i),MS}_{hh3,6,9,12,16} &= -\frac{2}{\sqrt{3}} \cdot \mathcal{O}^{(i),MA}_{ff1,4,7,10,14}, \nonumber\\
\mathcal{O}^{(i),MS}_{hh13,17,18} &= +\frac{2}{\sqrt{3}} \cdot \mathcal{O}^{(i),MA}_{ff13,17,18}.
\end{align}

There are further identities between the $MA$ three-quark operators:
\begin{align}
\mathcal{O}^{(i),MA}_{gg1,4,7,10,14} &= \mathcal{O}^{(i),MA}_{ff2,5,8,11,15}, \nonumber\\
\mathcal{O}^{(i),MA}_{gg2,5,8,11,15} &= \mathcal{O}^{(i),MA}_{ff1,4,7,10,14}, \nonumber\\
\mathcal{O}^{(i),MA}_{gg3,6,9,12,13,16,17,18} &= - \mathcal{O}^{(i),MA}_{ff3,6,9,12,13,16,17,18},
\end{align}
\begin{align}
\mathcal{O}^{(i),MA}_{hh1,4,7,10,14} &= \mathcal{O}^{(i),MA}_{ff2,5,8,11,15} - \mathcal{O}^{(i),MA}_{ff3,6,9,12,16}, \nonumber\\ 
\mathcal{O}^{(i),MA}_{hh2,5,8,11,15} &= \mathcal{O}^{(i),MA}_{ff2,5,8,11,15} - \mathcal{O}^{(i),MA}_{ff3,6,9,12,16}, \nonumber\\ 
\mathcal{O}^{(i),MA}_{hh3,6,9,12,13,16,17,18} &= 0.
\end{align}
Finally:
\begin{align}
\mathcal{O}^{(i),MA}_{ff18} &= \mathcal{O}^{(i),MA}_{ff17}.
\end{align}
Thus, as independent operators we may choose the set of $MA$ isospin operators with both covariant derivatives acting on the first quark $\mathcal{O}^{(i),MA}_{ff1,\dots,17}$.

In the case of the operators with the covariant derivatives acting on different quarks, we can express all of them by the $MA$ isospin combinations with derivatives acting on the second and third quark field. Then we arrive at similar equations as above:
\begin{align}
\mathcal{O}^{(i),MS}_{gh1,4,7,10,14} &= -\frac{1}{\sqrt{3}} \cdot \mathcal{O}^{(i),MA}_{gh1,4,7,10,14}, \nonumber\\ 
\mathcal{O}^{(i),MS}_{gh2,5,8,11,15} &= +\frac{1}{\sqrt{3}} \cdot \mathcal{O}^{(i),MA}_{gh2,5,8,11,15} - \frac{2}{\sqrt{3}} \cdot \mathcal{O}^{(i),MA}_{gh3,6,9,12,16}, \nonumber\\
\mathcal{O}^{(i),MS}_{gh3,6,9,12,16} &= -\frac{2}{\sqrt{3}} \cdot \mathcal{O}^{(i),MA}_{gh2,5,8,11,15} + \frac{1}{\sqrt{3}} \cdot \mathcal{O}^{(i),MA}_{gh3,6,9,12,16},\nonumber\\
\mathcal{O}^{(i),MS}_{gh13,17,18} &= -\frac{1}{\sqrt{3}} \cdot \mathcal{O}^{(i),MA}_{gh13,17,18},
\end{align}
\begin{align}
\mathcal{O}^{(i),MS}_{fh1,4,7,10,14} &= -\frac{1}{\sqrt{3}} \cdot \mathcal{O}^{(i),MA}_{gh2,5,8,11,15} + \frac{2}{\sqrt{3}} \cdot \mathcal{O}^{(i),MA}_{gh3,6,9,12,16}, \nonumber\\
\mathcal{O}^{(i),MS}_{fh2,5,8,11,15} &= +\frac{1}{\sqrt{3}} \cdot \mathcal{O}^{(i),MA}_{gh1,4,7,10,14}, \nonumber\\
\mathcal{O}^{(i),MS}_{fh3,6,9,12,16} &= -\frac{2}{\sqrt{3}} \cdot \mathcal{O}^{(i),MA}_{gh2,5,8,11,15} + \frac{1}{\sqrt{3}} \cdot \mathcal{O}^{(i),MA}_{gh3,6,9,12,16}, \nonumber\\
\mathcal{O}^{(i),MS}_{fh13,17,18} &= -\frac{1}{\sqrt{3}} \cdot \mathcal{O}^{(i),MA}_{gh13,17,18},
\end{align}
\begin{align}
\mathcal{O}^{(i),MS}_{fg1,4,7,10,14} &= -\frac{1}{\sqrt{3}} \cdot \mathcal{O}^{(i),MA}_{gh2,5,8,11,15} - \frac{1}{\sqrt{3}} \cdot \mathcal{O}^{(i),MA}_{gh3,6,9,12,16}, \nonumber\\
\mathcal{O}^{(i),MS}_{fg2,5,8,11,15} &= +\frac{1}{\sqrt{3}} \cdot \mathcal{O}^{(i),MA}_{gh2,5,8,11,15} + \frac{1}{\sqrt{3}} \cdot \mathcal{O}^{(i),MA}_{gh3,6,9,12,16}, \nonumber\\
\mathcal{O}^{(i),MS}_{fg3,6,9,12,16} &= -\frac{2}{\sqrt{3}} \cdot \mathcal{O}^{(i),MA}_{gh1,4,7,10,14}, \nonumber\\
\mathcal{O}^{(i),MS}_{fg13,17,18} &= +\frac{2}{\sqrt{3}} \cdot \mathcal{O}^{(i),MA}_{gh13,17,18},
\end{align}
\begin{align}
\mathcal{O}^{(i),MA}_{fh1,4,7,10,14} &= \mathcal{O}^{(i),MA}_{gh2,5,8,11,15}, \nonumber\\ 
\mathcal{O}^{(i),MA}_{fh2,5,8,11,15} &= \mathcal{O}^{(i),MA}_{gh1,4,7,10,14}, \nonumber\\ 
\mathcal{O}^{(i),MA}_{fh3,6,9,12,13,16,17,18} &= - \mathcal{O}^{(i),MA}_{gh3,6,9,12,13,16,17,18},
\end{align}
\begin{align}
\mathcal{O}^{(i),MA}_{fg1,4,7,10,14} &= \mathcal{O}^{(i),MA}_{gh2,5,8,11,15} - \mathcal{O}^{(i),MA}_{gh3,6,9,12,16}, \nonumber\\ 
\mathcal{O}^{(i),MA}_{fg2,5,8,11,15} &= \mathcal{O}^{(i),MA}_{gh2,5,8,11,15} - \mathcal{O}^{(i),MA}_{gh3,6,9,12,16}, \nonumber\\ 
\mathcal{O}^{(i),MA}_{fg3,6,9,12,13,16,17,18} &= 0.
\end{align}
Finally:
\begin{align}
\mathcal{O}^{(i),MA}_{gh18} &= \mathcal{O}^{(i),MA}_{gh17} .
\end{align}
Therefore we choose as independent operators the reduced set of $MA$ isospin operators with the covariant derivatives acting on the second and third quark: $\mathcal{O}^{(i),MA}_{gh1,\dots,17}$.

%
%

\end{document}